\documentclass[letterpaper,twocolumn,10pt]{article}
\usepackage{usenix}
\usepackage{hyperref}
\usepackage[hyphenbreaks]{breakurl}
\usepackage{amssymb,amsfonts}
\usepackage{balance}
\usepackage{amsmath, amsthm}
\usepackage{mathtools, nccmath}
\usepackage{url}
\usepackage{color}
\usepackage{caption}
\usepackage{diagbox}
\usepackage{subfigure}
\usepackage{multirow}
\usepackage{booktabs}
\usepackage{epsfig,endnotes}
\usepackage{algorithm,algpseudocode}
\usepackage{enumitem}
\usepackage{algpseudocode}
\usepackage{times}

\newcommand{\RNum}[1]{\uppercase\expandafter{\romannumeral #1\relax}}

\newtheorem{theorem}{Theorem}

\DeclareMathOperator*{\argmin}{argmin}

\newcommand{\para}[1]{{\vspace{2pt} \noindent \textbf{#1}
    \hspace{6pt}}}

	\definecolor{applegreen}{rgb}{0.55, 0.71, 0.0}

\newcommand{\abedit}[1]{{\color{black} #1}}
\newcommand{\shawnmr}[1]{{\color{black} #1}}
\newcommand{\shawn}[1]{{\color{black} #1}}
\newcommand{\minor}[1]{{\color{black} #1}}

\newcommand{\htedit}[1]{{\color{black} #1}}

\newcommand{\ssedit}[1]{{\color{black} #1}}

\newcommand{\etal}{{\em et al.\ }}
\newcommand{\eg}{{\em e.g.,\ }}
\newcommand{\ie}{{\em i.e.,\ }}

\newcommand{\secspace}{\vspace{-0.08in}}

\newcommand{\ad}[1]{{$\mathcal{A}$}}
\newcommand{\service}[1]{{$\mathcal{S}$}}
\newcommand{\mcD}{\mathcal{D}}

\newenvironment{packed_itemize}{
\begin{list}{\labelitemi}{\leftmargin=0.5em}
  \setlength{\itemsep}{1pt}
  \setlength{\parskip}{0pt}
  \setlength{\parsep}{0pt}
  \setlength{\headsep}{0pt}
  \setlength{\topskip}{0pt}
  \setlength{\topmargin}{0pt}
  \setlength{\topsep}{0pt}
  \setlength{\partopsep}{0pt}
}{\end{list}}

\begin{document}

\title{Poison Forensics: Traceback of Data Poisoning Attacks in Neural Networks}

\author{Shawn Shan, Arjun Nitin Bhagoji, Haitao Zheng, Ben Y. Zhao\\
{\em Computer Science, University of Chicago}\\
{\em \{shawnshan, abhagoji, htzheng, ravenben\}@cs.uchicago.edu}}


\maketitle



\begin{abstract}
  In adversarial machine learning, new defenses against attacks on deep
  learning systems are routinely broken soon after their release by more
  powerful attacks. In this context, {\em forensic tools} can offer a valuable
  complement to existing defenses, by tracing back a successful attack to its
  root cause, and offering a path forward for mitigation to prevent similar
  attacks in the future.

  In this paper, we describe our efforts in developing a forensic traceback
  tool for poison attacks on deep neural networks. We propose a novel
  iterative clustering and pruning solution that trims ``innocent'' training
  samples, until all that remains is the set of poisoned data responsible for
  the attack. Our method clusters training samples based on their impact on
  model parameters, then uses an efficient data unlearning method to prune
  innocent clusters.  We empirically demonstrate the efficacy of our system
  on three types of dirty-label (backdoor) poison attacks and three types of
  clean-label poison attacks, across domains of computer vision and malware
  classification. Our system achieves over $98.4\%$ precision and $96.8\%$ recall
  across all attacks. We also show that our system is robust against four
  anti-forensics measures specifically designed to attack it.
  
\end{abstract}

\secspace
\section{Introduction}
\label{sec:intro}
\vspace{-0.06in}

For external facing systems in real world settings, few if any security
measures can offer full protection against all attacks. In practice, digital
forensics and incident response (DFIR) provide a complementary security tool
that focuses on using post-attack evidence to trace back a successful attack
to its \emph{root cause}. \htedit{For packet routing on the wide-area
  Internet}, for example, forensic IP 
traceback tools can identify the true source of a Denial of Service (DoS)
attack. Not only can forensic tools help operators identify (and hopefully
patch) vulnerabilities responsible for successful attacks, but strong
forensics can provide a strong deterrent against future attackers by
threatening them with post-attack identification.

Such an approach would be particularly attractive in the context of
attacks against deep learning systems, where new defenses are routinely
broken soon after their release by more powerful 
attacks~\cite{wenger2021backdoor,severi2021explanation,carlini2016defensive,athalye2018obfuscated,carlini2017adversarial}. Consider
for example ``poisoning attacks,'' a threat that arises from the reliance of
ML trainers and operators on external data sources, either purchasing data
from or outsourcing data collection to third
parties~\cite{papernot2018marauder}. An attacker can inject manipulated
training data into the training data pipeline, thus causing the resulting
model to produce targeted misclassification on specific inputs. Recent
advances in poisoning attacks have made them more
powerful~\cite{aghakhani2020bullseye,liu2018trojaning}, more
realistic~\cite{severi2021explanation,wenger2021backdoor,yao2019latent}, and
more stealthy~\cite{liao2018backdoor,bagdasaryan2020blind}.  In a recent
survey, industrial practitioners ranked data poisoning attacks as the
most worrisome threat to industry machine learning
systems~\cite{kumar2020adversarial}.

For data poisoning attacks, effective forensics would add a valuable
complement to existing defenses, by helping to identify {\em which} training
samples led to the misclassification behavior used in the attack. We call
this the ``poison traceback problem.'' Starting with evidence of the attack
(an input sample that triggers the misclassification), a forensic tool would
seek to identify a particular subset of training data responsible for
corrupting the model with the observed misclassification behavior.  Combined
with metadata or logs that track the provenance of training data, this
enables practitioners to identify either the source of the poison data, or a
vulnerability in the data pipeline where the poison data was inserted. Either
result leads to direct mitigation steps (e.g. removing an unreliable data
vendor or securing a breached server on the training data pipeline) that
would patch the pipeline and improve robustness to similar attacks in the
future.


Several factors make the poison traceback problem quite challenging in
practice. First, today's deep learning models employ large complex
architectures that do not easily lend themselves to explainability. Specific
behaviors do not localize themselves to specific neurons as once
speculated. Second, the effects of poisoning attacks generally require
training on a {\em group} of poisoned data, and a subset of the poisoned
training data is unlikely to produce the same behavior.  Thus a brute force
search for the subset of poisoned training data would involve testing an
exponential number of sample combinations from a large training
corpus. Finally, a poison traceback tool produces evidence that can lead to the
identification of parties responsible for an attack. Thus these tools
must have very high precision, since false positives could lead to false
accusations and negative consequences.

In this paper, we introduce the poison traceback problem, and propose \ssedit{the first}
solution that accurately identifies poisoned training data responsible for an
observed attack. Our solution utilizes an iterative clustering and pruning
algorithm. At each step, it groups training samples into clusters based on
their impact on model parameters, then identifies benign clusters using an
efficient data unlearning algorithm. As benign clusters are pruned away, the
algorithm converges on a minimal set of training samples responsible for
inducing the observed misclassification behavior. In detailed experiments
covering a variety of tasks/datasets and attacks, our approach produces
highly accurate (high precision and recall) identification of poison data for
both dirty-label and clean-label poison attacks.


This paper makes the following contributions to the forensics of poison
attacks:

\begin{packed_itemize}\vspace{-0.05in}
\item We define forensics in the context of data poisoning attacks and design
  a forensics system that effectively traces back misclassification events to poison
  training data responsible.
\item We empirically demonstrate the effectiveness of our system on three
  types of dirty-label poison attacks and three types of clean-label
  poison attacks, across two domains of computer vision and malware classification.
  Our system achieves over $98.4\%$ precision and $96.8\%$ recall on
  identifying poison training data,
\item \shawnmr{We test our system against 4 alternative forensic designs
    adapted from prior defenses, and show our system consistently succeeds on
    attacks where alternatives fall short.}
\item We consider potential anti-forensics techniques that can be used to
  evade our system. We test our system and show that it is robust against 6
  adaptive attacks specifically designed to overcome this forensic system.
  \vspace{-0.08in}
\end{packed_itemize}

To the best of our knowledge, this is the first work to explore a forensics
approach to address data poisoning attacks on deep learning systems. 
This is a significant departure from existing works that focus entirely on
attack prevention. Given our initial results, we believe poison 
traceback is a promising direction worthy of further exploration.


\secspace
\section{Background and Related Work}
\label{sec:back}
\vspace{-0.05in}

In this section, we present the background and related work on data poisoning and digital forensics.

\secspace
\subsection{Data Poisoning}

In data poisoning attacks, the attacker gains access to the training data pipeline of the 
victim ML system, \eg via a malicious data provider, and injects a set of poison data into the training dataset. The poison data causes the victim's model to have 
certain vulnerabilities, \ie misclassifying certain inputs targeted by the attacker. 

\para{Data Poisoning Attacks. } We can divide existing poison attacks into two categories 
based on their attack assumptions: dirty-label attacks where attacker can modify 
both the data and their semantic labels, and clean-label attacks where attacker can only 
modify the data. Dirty-labels attacks~\cite{gu2017badnets,liu2018trojaning,
wenger2021backdoor}, often called as backdoor attacks, seek to inject a \emph{trigger} into 
the victim model. A trigger is a unique input signal (\eg a yellow sticker on an image, a 
trigger word in a sentence) that once present can lead the victim model to misclassify any 
inputs to a target label selected by the attacker (\eg the presence of yellow sticker leads the  
model to classify stop signs as speed limits~\cite{gu2017badnets}). 

Clean-label attacks further divide into clean-label backdoor attacks and clean-label 
triggerless attacks. Clean-label backdoor attacks~\cite{severi2021explanation,turner2018clean,saha2020hidden} are similar to dirty-label backdoor attacks except that attacker cannot modify the 
label of the poison data. Clean-label triggerless attack aims to misclassify a single 
unmodified test data. Shafahi \etal~\cite{shafahi2018poison} proposed the first clean-label 
triggerless attack where an attacker injects poison data to disrupt the feature region of the 
targeted data. Several proposals~\cite{aghakhani2020bullseye,zhu2019transferable} significantly 
improve the performance of clean-label attacks by positioning poison data on a convex 
polytope around the target data. These clean-label attacks only perform well when the 
victim model's feature space is known, \ie assuming the victim uses transfer 
learning and the attacker has white-box access to the pretrained model's parameters. A 
recent attack, WitchBrew~\cite{geiping2020witches}, targets the from-scratch training 
scenario leveraging gradient alignment of poison and target data. 

\para{Data Poisoning Defenses. } A large body of research seeks to defend against poison 
attacks. Robust training defenses modify the training of neural networks to be 
resilient against data poisoning. Existing robust training defenses leverage ensemble 
training~\cite{jia2020intrinsic}, kNN majority voting~\cite{jia2020certified}, adversarial 
training~\cite{geiping2021doesn}, random smoothing~\cite{wang2020certifying}, and data 
augmentation~\cite{borgnia2021strong}. Other defenses try to diagnose and patch an already 
poisoned model. Neural Cleanse~\cite{wang2019neural} assumes backdoor triggers are small 
input signals and reverse engineers the injected backdoor trigger. 
Fine-Pruning and STRIP assume neurons related to poison are not activated by benign data, 
and thus remove unused neurons~\cite{liu2018fine,gao2019strip}. SPECTRE~\cite{
hayase2021defense} assumes a Gaussian distribution of benign feature representations, and 
filters out anomalous inference queries. 


Still, defending against poison attacks remains a challenging problem, mainly because the injected 
vulnerability is hidden and has not been activated at defense time. Thus, existing defenses 
examine the training data or various behaviors of the model to identify anomalous signals 
that might be malicious. While existing defenses have shown promising signs by preventing many poison attacks, stronger and adaptive attackers are able to bypass existing 
defenses~\cite{wenger2021backdoor,yao2019latent,severi2021explanation,bagdasaryan2020blind, 
  schuster2021you}.

\begin{figure*}[t]
  \centering
  \includegraphics[width=0.9\linewidth]{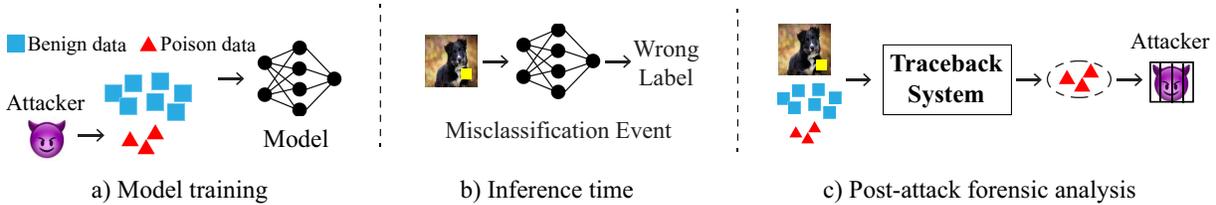}
  \vspace{-0.1in}
  \caption{The general scenario for our trackback system.  a) the attacker poisoned the training data to inject vulnerability into the model;  b) at run-time, the attacker submits an attack input to cause a misclassification event;  c) our traceback system inspects the misclassification event to identify its root cause. }
  \label{fig:forensic_problem}
  \vspace{-0.1in}

\end{figure*}

\secspace
\subsection{Background on Digital Forensics}
\label{sec:back_forensics}

First introduced in the 1970s, digital forensics has been a major and growing area of 
cybersecurity. Digital forensics seeks to trace the \emph{source} of a cyberattack that has 
already happened leveraging traces that the attacker left in the victim system. 



``Attack incidents'' trigger the forensic analysis, \ie when the system administrator discovers 
a cyberattack after some catastrophic events have happened (\eg web servers overloaded with 
dummy requests, machine takeover, or sensitive data appearing in the dark web). Then, a 
forensics system is assigned to investigate the source of the attack. 
Forensic analysis often starts with evidence collection from the logs of the victim system. Then 
the system connects these pieces of evidence using their casual links to form a causal graph, and 
identifies the root cause of the attack by tracing through the casual 
graph starting from the attack incident. 

\para{Benefits of Forensics.} Successful forensics can lead to prosecution
of the perpetrator, stopping the attack from the source, and offering
insights to build more secure systems. Forensics can even break the arms
race between attackers and defenders, since an attacker faces a much higher
cost of iterating with a forensics system, \ie the attacker is held
accountable as long as the forensics system succeeds once. Consequently, the risk of
being caught acts as a strong deterrent to discourage any attackers from
launching the attack in the first place.


\para{Forensics vs. Defenses.} Forensics is a \emph{complementary} approach
to defenses (or security through prevention).  While there is a significant
amount of prior works focusing on defenses against adversarial attacks,
history (in both machine learning security and multiple other security areas)
shows that no defense is perfect in practice, and attackers will find ways to
circumvent even strong defenses. Forensics addresses the incident response of
successful attacks by tracing back to the root causes. Modern security
systems leverage both defenses and forensics to achieve maximum security.

\shawnmr{
The same dynamic holds true in the context of poison attacks on neural
networks. For example, a defense against backdoors that identifies poison
training data can be circumvented by an attacker who breaches the server
after the defense has been applied, but prior to model training. 
}

\para{Existing Digital Forensics Research. } Forensics has been widely studied in security 
community to solve a wide variety of security problems, \eg tracing the source IP of DDoS 
attacks~\cite{savage2000practical,dean2002algebraic,song2001advanced}, origins of intrusion~
\cite{king2003backtracking,yin2007panorama}, and the cause of advance persistent threats
(APTs)~\cite{xu2016high,yu2021alchemist,fei2021seal}. Existing research addresses many 
technical challenges of forensics. \cite{bates2015trustworthy,karande2017sgx,eo2015phase,
lazaridis2016evaluation} seek to secure the integrity of traces left by the attacker 
against potential tampering. \cite{xie2012hybrid,lee2013loggc,ma2018kernel,ding2021elise} 
reduce the large storage overhead of logging while preserving enough 
information. \cite{yu2021alchemist,ma2016protracer,king2003backtracking} address the 
dependency-explosion problem where a forensics system cannot narrow down the true cause of 
the attack. Another line of research focuses on post-forensics, \ie after the root cause is 
identified. The post-forensics system can prosecute the attacker in court by generating 
causal proof~\cite{garrie2014digital,efendi2019management} and fingerprint the attack to 
prevent similar attacks in the future~\cite{newsome2005dynamic,kaczmarczyck2020spotlight,
perdisci2010behavioral}.

\secspace
\section{Traceback on Data Poisoning Attacks}
\label{sec:define}

In this paper, we consider the task of applying forensics to uncover the presence of data poisoning attacks on deep neural network (DNN) models. Given a misclassification event at test time, we seek to identify the set of poisoned training data that resulted in the misclassification.


\para{Example Scenario.} Figure~\ref{fig:forensic_problem} illustrates the
general scenario for post-attack forensic analysis.  One or more attackers
find a way to access the 
training data pipeline\footnote{The training data pipeline often
  includes multiple layers of data collectors, labelers, and brokers.},
and inject poison training data to introduce a specific vulnerability
into the DNN model (Figure~\ref{fig:forensic_problem}(a)). Once the corrupted
model is deployed, the attacker submits a carefully crafted input that
exploits the vulnerability to produce a misclassified result. When the administrator discovers
this misclassification event (possibly after downstream events), they want
to identify the root cause or entity responsible
(Figure~\ref{fig:forensic_problem}(b)).  Information is sent to the
traceback system, including the input that caused the
misclassification, the DNN model, and its training data.  The traceback
system then identifies the poison training data responsible for the misclassification event
(Figure~\ref{fig:forensic_problem}(c)).

Here, we define our threat model, identify key goals and challenges of a
forensic traceback system for poisoning attacks, and highlight our key
intuition for our solution.  We describe the
details of our traceback system in \S\ref{sec:method}.

\para{Terminology.}  We use the following terminology:
\begin{packed_itemize} \vspace{-0.05in}
\item \textbf{Data poisoning attack}: the injection of poisoned training data that embeds vulnerability into the victim model. 
\item \textbf{Misclassification event}: an input ($x_a$) that the
  model misclassified, and the corresponding misclassified label ($y_a$). 

\end{packed_itemize}\vspace{-0.08in}

\secspace
\subsection{Threat Model}\vspace{-0.05in}
We first describe our threat model and assumptions of the 
attacker and the traceback system.

\para{Data Poisoning Attacker. } We adopt common assumptions made by existing
work on poisoning attacks and defenses. We assume the attacker: 

\begin{packed_itemize} \vspace{-0.05in}
\item can modify any portion of their controlled training data;

\item can poison at most half of the entire training dataset;

\item has no access to other parts of the model training pipeline, including
  the final model parameters of the trained model\footnote{There exists a few
    parameter-space poison
    attack~\cite{hong2021handcrafted,shumailov2021manipulating} and we
    consider them outside of our threat model since these attacks require
    additional access to the victim's training pipeline. };
\item is aware of the existence of a potential traceback system and can adopt
  anti-forensics techniques to evade traceback (more details in
  \S\ref{sec:counter}); 
\item at inference time, submits an attack input that utilizes the injected
  vulnerability to cause model misclassification.
\end{packed_itemize}
\vspace{-0.08in}

\para{Traceback System. } We assume the traceback system is deployed by the
model owner or a trusted third party, and thus has full access to the
following resources:
\begin{packed_itemize} \vspace{-0.05in}
\item the DNN model (its parameters and architecture), the model training
  pipeline and the training data;
\item information on the misclassification event, \ie the exact attack input
  $x_a$ and misclassification output $y_a$.
\end{packed_itemize} \vspace{-0.08in}
We do not assume the traceback system has any access to
information on other attacks (beyond the current misclassification event), and make no assumption about
types or parameters of the poisoning attacks. 
\minor{
Note that the traceback system has full access to the training dataset, unlike
assumptions made by some existing defenses, which prevent poison attacks without leveraging the 
full training dataset~\cite{wang2019neural}. 
}

\htedit{We note that in practice, a misclassification event may also
  arise from low model accuracy or from an evasion attack. Either can
  cause misclassification without poisoning/modifying training
  data. In \S\ref{sec:discussion}, we discuss how our forensic tool can also be
  used to determine if a misclassification was caused by a
  poisoning attack. 
  A comprehensive study of robust recognition of non-poison misclassification
  events is beyond the scope of this paper. In the rest of the paper, we only
  limit ourselves to data poisoning attacks.}


\secspace
\subsection{Design Requirements and Challenges} 

To identify what/who is responsible for the misclassification event, a
practical DNN traceback system should meet the following requirements:
\begin{packed_itemize} \vspace{-0.05in}
\item {\bf High precision} -- In forensics, false positives can lead to false
  accusations, and thus must be minimized. Under our problem context, this
  means that for any misclassification event caused by a specific
  poisoning attack $\mathbb{A}$, traceback should identify only those
  poisoned training data injected to implement $\mathbb{A}$ but not
  others.

\item {\bf High recall} -- Recall measures the percentage of poison 
  training data responsible for the misclassification event that are identified by the
  traceback system.  Achieving a high recall rate is crucial for identifying
  all the attack parties, especially when multiple parties worked together to
  inject poison data in order to train a vulnerability into the
  model.

\item {\bf Generalizability} -- An effective traceback system should address
  a wide range of poisoning attacks against DNN models, without requiring
  knowledge of the attack type or parameters (\eg the amount of
  poison training data).


\end{packed_itemize} \vspace{-0.05in}

We further note two {\em non-goals} of our system. The first is
{\em attack scope}.  The goal of the traceback system is to respond to a specific, observed
attack. In a scenario where one or more attackers have performed multiple,
independent poisoning attacks on the same model, the traceback system focuses
on identifying the poison data that caused the observed
misclassification event.  The second is on {\em
  computational latency}.  Unlike real-time attack detection tools, forensic
traceback is a post-attack operation and does not face strict latency
requirements. This is a common assumption for digital forensics~\cite{casey2009handbook,gogolin2021digital}.


\para{Potential Solutions and Key Challenges.} Traditional digital forensics
traces the root cause of an attack by building causal graphs using causal
links among system
events~\cite{yu2021alchemist,ma2016protracer,king2003backtracking}.  In our
problem context, DNN model training allows each individual training sample to
potentially contribute to the final model parameters, and by extension, the
misclassification result. This is commonly known in forensics as the {\em
  dependency explosion} problem, and combined with the large size of training
data (\eg millions of training samples) makes conventional causal graph
analysis intractable.

The key challenge facing any DNN forensic system is how to efficiently
connect a (mis)classification result to specific samples in the training data.  For
non-DNN, linear ML models, existing works use the well-known {\em
  influence function}~\cite{koh2017understanding} to estimate the
contribution of each training data point towards a classification result,
leveraging the first-order Taylor's approximation.  However, recent
work~\cite{basu2020influence} showed that when applied to DNNs,  the 
influence function produces poor performance and requires costly computation
of second-order derivatives.  We confirmed these observations 
experimentally. Using the influence function to traceback BadNet poisoning
attacks on a CIFAR10 model (details in \S\ref{sec:eval}) achieved less
than 69\% precision and recall.  When testing it on models trained on the larger
ImageNet, our influence function computation timed out after running
15 days on 4 NVidia TitanX GPUs.

\shawnmr{
Another alternative is to adapt existing poison defenses into forensic tools
for use after an attack has been detected. While adapting defense techniques
as forensics is itself slightly paradoxical (waterproof defenses would
obviate the very need for forensics), we can nonetheless test to see if such
techniques can be effective after an attack. 
Later in \S\ref{subsec:compare}, we adapt four defenses (Spectral Signature,
Neural Cleanse, Deep K-NN, and $L_2$-Norm) into potential forensic tools, and
compare them with our traceback system. Some of the adapted systems
have success against simple attacks, but all of them
fail on stronger poison attacks.}

\para{Poisoning as a group effect.} In current work on poisoning attacks,
attack success relies on a critical mass of poison samples in the
training set~\cite{gu2017badnets,severi2021explanation,shan2020fawkes}.  While a sufficiently large set of poison
training data can shape model behavior and inject vulnerabilities, the
contribution of each individual data sample is \htedit{less and much
harder to quantify.} This explains the poor performance of the influence function when
applied to smaller models such as CIFAR10 (see above). It motivates us
to design a solution to search for groups of poison training data, not single samples.

\secspace
\subsection{Design Intuition}
\label{sec:intuition}

Instead of designing the traceback system to explicitly target individual
training data samples, our intuition is to inspect training data in groups, and map the traceback problem to a {\em set searching}
problem.

\para{Set searching by iterative clustering and pruning.}  We propose to
search for sets of training samples responsible for an observed
misclassification event, by
iteratively pruning groups of training data we can identify as \ssedit{{\em
  innocent}} to the attack.  Starting with the full training data set, we
progressively identify and prune clusters of \ssedit{innocent} training samples until only those {\em
  responsible} for the misclassification event are left.  In each iteration, we only need to
identify clusters of training data that {\em do not} contain any
poison training 
data required to make the misclassification event successful.  As such, our traceback design
only needs a ``binary'' measure of event responsibility, which is much
easier to compute than the actual contribution of any training data
samples to the
attack.

\para{A binary measure of event responsibility.}  We propose a binary
measure of event responsibility, which connects a misclassification event with
the model training data.  Here our hypothesis is that since data poisoning
attacks focus on making the model learn new behavior that is different from
those offered by the benign (or innocent) data,  \htedit{the attack
  confidence level} should not degrade if some portion of the innocent data
is \htedit{\emph{not} used for model training}.  \htedit{In this work, we
  propose to use this condition to determine whether a cluster of training data contains
only innocent data not responsible for the misclassification. }

We formally define this condition as follows. Let the model's full training
dataset $D$ be divided into two distinct subsets: $D_1$ and
$D\setminus D_1$. Let
$\mathcal{F}$ be the DNN model trained on $D$ and $\mathcal{F}^-$ be the 
model trained on $D\setminus D_1$.  Let  $(x_a,y_a)$ represent the
misclassification event.  We use 
$\ell (\mathcal{F}(x_a), y_a)$ and $\ell (\mathcal{F}^-(x_a), y_a)$ to 
indirectly \htedit{compare the confidence level of $(x_a,y_a)$} 
on the two DNN models, where
$\ell(.)$ is the cross-entropy loss function.  Specifically, if removing $D_1$ from the model training data
does not increase the attack confidence level, \ie 
\begin{equation}
  \ell (\mathcal{F}(x_a), y_a) \geq \ell (\mathcal{F}^-(x_a), y_a),
  \label{eq:binarymetric}
\end{equation}
\abedit{then $D_1$ is \emph{less responsible} for the misclassification event $(x_a,y_a)$ than $D \setminus D_1$. This is in the sense that $D \setminus D_1$ has a ratio of benign to poison data that is more skewed towards poison than $D$, allowing us to use this measure of event responsibility to iteratively determine the subset of poison data. We note that in practice, we are able to use clustering to find splits such that $D_1$ does not contain \emph{any} poison data.} Our proposed measure {\em only} examines the attack confidence level, and does
not consider the model's normal classification accuracy. 


\abedit{The proposed binary measure of event responsibility (Eq. \ref{eq:binarymetric}) is supported by our theoretical analysis on how removing a portion of the training data affects attack performance.  For brevity, we present the analysis in Appendix \ref{appsec: theorem_proof}.  }

\secspace

\section{Detailed Poison Traceback Design}
\label{sec:method}
This section presents the detailed design of our traceback system.  We start from a high-level overview, followed by the detailed description of the two key components (clustering and pruning), \htedit{which run iteratively to identify the set of poison training data responsible for the misclassification event.}




\secspace
\subsection{High-level Overview}
\label{subsect:designoverview}
In a nutshell, our traceback system implements an iterative clustering and pruning process, which progressively identifies sets of innocent training data that are not responsible for the observed misclassification event. This ends when only the poison data responsible for the misclassification event is left and thus identified.   Doing so requires two key operations: clustering and pruning. 

\para{(1)  Clustering unmarked training data.} The clustering component divides the unmarked training data into clusters, based on how they affect the model parameters (details in \S\ref{subsec:cluster}).  Here the goal is to progressively separate innocent data from poison data, so that we can identify, mark and prune an innocent cluster (using the pruning component).   

\para{(2) Identifying and pruning innocent clusters.}  This pruning component examines the unmarked data clusters,  applies the binary metric defined by Eq. (\ref{eq:binarymetric})  to identify an {\em innocent} cluster, if any, that is not responsible for the misclassification event.  The identified cluster is marked and thus excluded from the next clustering operation, \ie pruned out.  We note that pruning does not affect the computation of Eq. (\ref{eq:binarymetric}), where $D$ is always the original full training dataset and $D_1$ is a cluster to be examined. The detailed pruning design is in \S\ref{subsec:prune}.

\begin{figure}[h]
  \centering
  \includegraphics[width=0.85\linewidth]{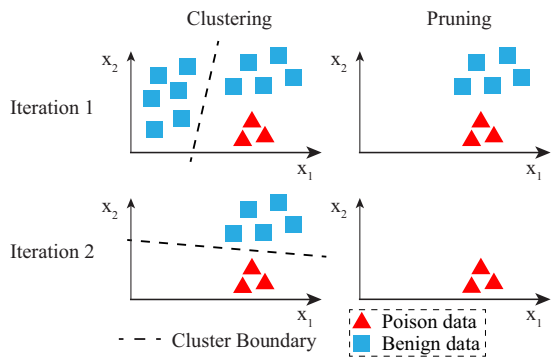}
  \caption{An illustration of our poison traceback process that completes in two iterations,  visualized on a simplified 2D space representing the training data.} 
  \label{fig:feature_illustration}
  \vspace{-0.2in}
\end{figure}

\para{An illustrative example.} Figure~\ref{fig:feature_illustration} shows an example traceback process that completes in two iterations, visualized in a simplified 2D projection of the training data. The traceback starts from the full set of training data as unmarked, including both innocent (blue) and poison (red) data. In each iteration, the left figure shows the collection of unmarked training data to be clustered and the resulting cluster boundary that divides them into two clusters;  the right figure shows the result of pruning where the innocent cluster is removed. At the end of iteration 2, only the set of poison data responsible for the attack is left as unmarked. Upon detecting that the unmarked data cannot be further divided or pruned, the process ends and the unmarked data are declared as the training data responsible for the misclassification event. 


\secspace
\subsection{Clustering Unmarked Training Data}
\label{subsec:cluster}
We now describe the detailed clustering design, which seeks to progressively separate benign and poison training data into different clusters.  The key challenge here is that while innocent and poison data are different by design (in order to inject different behaviors into the model), it is highly challenging to accurately characterize and measure such differences.  This is also why it has been hard to design defense mechanisms that can effectively identify and remove poison training data. 


Instead, our clustering design first maps these training data into a new space, focusing on  ``amplifying'' the separation between innocent and poison data rather than identifying them. Operating on this new space, each clustering operation will produce two clusters,  ideally one containing only innocent data and the other containing a mixture of innocent and poison data. Given these two clusters, our pruning component (\S\ref{subsec:prune})  will replay the misclassification event to identify the innocent cluster. In the next iteration,  we will run clustering only on the mixed cluster to ``extract'' more innocent data.  After a few iterations, the innocent and poison data become fully separated.

Therefore, our clustering design includes 1) data mapping to ``amplify'' the distance between innocent and poison data,  and 2) a high performance clustering method to generate the clusters, which we discuss below. 


\para{Data mapping. }  We map the data by estimating how a training sample $x$ affects the final model 
parameters. This is measured by the change of model parameters when $x$ is absent from the training dataset, \ie comparing the final model parameters when trained on $D$ and $D\setminus x$, where $D$ is the full training 
dataset. \minor{Unlearning benign or poison data results in \textit{different} effect on model parameters. 
Unlearning of poison samples shifts the model closer to an optimal location in parameter space, where poisoning is ineffective, while unlearning benign samples shifts the model towards its initial randomly initialized state, since if all benign samples are effectively unlearned, the model will have no predictive power. 
} A naive implementation would retrain the model on $D\setminus x$, leading to unnecessary computational 
overhead and stochasticity from training. Instead, inspired by the concept of {\em unlearning}, we propose
estimating the parameter change using a gradient computation. \minor{The gradient of the parameters with respect to a given data point with a specified loss function is a well-known method to characterize its impact on the model~\cite{koh2017understanding}. Intuitively, data with similar gradients will have a similar impact on the model. Thus, our data mapping for training data point $x$ is: }

\vspace{-0.05in}
\begin{equation} 
  \nabla_{\theta} \: \ell (\mathcal{F}(x), NULL)
  \label{eq:reverse-gradient}
\end{equation}
\vspace{-0.22in}

where $\mathcal{F}$ is the original model (trained on $D$), $\theta$ is the parameter set of $\mathcal{F}$'s classification layer, $\ell$ is the 
cross-entropy loss, and $NULL$ is a new ``no knowledge learned'' label to represent the effect of not learning from $x$.  We implement $NULL$ as an equal probability output, which has been used by existing works to label out-of-distribution (OOD) samples~\cite{vyas2018out}.   

\minor{

We note that our data mapping method is one of the many ways to represent 
data for analysis in poison setting. Other mapping methods exist in 
the literature~\cite{chen2018detecting,carlini2021poisoning}. We leave a systematic study of 
the optimaility of data mappings and designing better performing mapping to future work. 

}

\para{Clustering heuristics. } To handle large training data sets,  we use Mini-Batch K-means~\cite{sculley2010web}, a scalable variant of the K-means clustering algorithm. As the name suggests, it first runs K-means on multiple smaller batches of the dataset and then aggregates their results.  This allows us to distribute the computation across multiple servers, achieving orders of magnitude speedup without degrading the clustering quality~\cite{sculley2010web}.  As discussed earlier,  we configure the clustering system to produce two clusters per iteration.


\secspace
\subsection{Pruning the Innocent Cluster}
\label{subsec:prune}
Given the two clusters, the pruning component will identify \abedit{which of the two clusters is less responsible, if any, for the misclassification event $(x_a,y_a)$. } This is done by using $(x_a,y_a)$ to evaluate each cluster's responsibility to the event --  if a cluster meets the condition in  Eq.~(\ref{eq:binarymetric}),  it is marked as innocent and excluded in the next clustering iteration. \abedit{In practice, we find that the less responsible cluster always contains only benign data, due to the clear separation induced by our mapping in \S \ref{subsec:cluster}, resulting in rapid identification of the poison data.}


The design challenge facing this component is how to efficiently evaluate the condition defined by  Eq. (\ref{eq:binarymetric}), especially  $\ell (\mathcal{F}^-(x_a), y_a)$. Given a cluster ($D_1$), $\mathcal{F}^-$ refers to the DNN model trained on $D\setminus D_1$.  One can certainly compute $\ell (\mathcal{F}^-(x_a), y_a)$ by training $\mathcal{F}^-$ from scratch, which is often expensive in practice.  Again, inspired by the concept of `unlearning', we propose producing $\mathcal{F}^-$ by unlearning $D_1$ from the original model $\mathcal{F}$.




\para{Existing unlearning for provable privacy protection. } Unlearning has been 
explored in the  context of privacy (\eg~\cite{bourtoule2021machine,guo2019certified,
  neel2021descent}) where a person protected by privacy laws can request their data be removed from a trained ML model.  Thus, existing unlearning methods focus on achieving a strong, provable privacy guarantee at the cost of high computation complexity. Furthermore, their performance degrades rapidly as the number of data points to be unlearned increases~\cite{bourtoule2021machine,guo2019certified}, making them unsuitable for our traceback system.


\para{Proposed: functional unlearning for traceback. } Instead, we propose approximating $\mathcal{F}^-$ by {\em functionally unlearning} a cluster $D_1$ from the original model $\mathcal{F}$.  Specifically, we fine-tune  $\mathcal{F}$ to minimize the cross-entropy loss between $D_1$ and the NULL label (\ie no knowledge learned)  discussed in \S\ref{subsec:cluster} while maintaining a low cross-entropy loss on the rest of the training data ($D\setminus D_1$) like the original model. This fine-tuning operation is guided by 

\begin{equation}
    \label{eq:unlearning}
      \underset{\boldsymbol{\theta}}{\text{min}} \left(\sum_{(x,y)\in D_1}\ell(\mathcal{F}(x), NULL) + \sum_{(x,y)\in D\setminus D_1}\ell (\mathcal{F}(x), y)\right) 
      \end{equation}

\vspace{-0.1in}

where $(x,y)\in D\setminus D_1$ represents the training data instance (input $x$ and its label $y$). We solve the above 
optimization using stochastic gradient descent (SGD) with the same hyperparameters as the 
original model training, and use the fine-tuned version of $\mathcal{F}$ as $\mathcal{F}^-$. We then
compute $\ell (\mathcal{F}^-(x_a), y_a)$ using the misclassification event $(x_a,y_a)$ and verify the condition 
defined by Eq. (\ref{eq:binarymetric}).

\secspace
\section{Overview of Evaluation}

\htedit{Using a variety of tasks/datasets, we evaluate our
traceback system against $6$ different poison attacks ($3$ dirty-label and $3$ clean-label 
attacks) and $4$ anti-forensic countermeasures.  We
outline these experiments and a
preview of our findings. }


\para{\RNum{1}. Traceback of dirty-label poison attacks
  (\S\ref{sec:eval}).} Using $4$ image classification datasets, we 
test against $3$ state-of-the-art dirty-label attacks, \htedit{including an attack without any known effective defense.} 
Our traceback system achieves $\geq 98.9\%$ 
precision and $\geq 97.1\%$ recall in identifying poison data. 

\para{\RNum{2}. Traceback of clean-label poison attack
  (\S\ref{sec:clean-label}).} For both image classification and
malware classification datasets, we 
test against $3$ state-of-the-art clean-label attacks, 
including an attack without any known effective defense. \htedit{Our traceback
system achieves $\geq 98.4\%$ precision and $\geq 96.8\%$ recall.} 


\para{\RNum{3}. Robustness against anti-forensic countermeasures 
  (\S\ref{sec:counter}).} We consider \htedit{$4$ potential} 
countermeasures that a resourceful attacker can deploy to bypass the 
traceback. Results show that our system is robust against
all four. \htedit{Across all the 
experiments, the most effective countermeasure 
reduces the traceback precision and recall by less than $4\%$. }

\secspace
\section{Evaluation on Dirty-label Attacks}
\label{sec:eval}
\vspace{-0.06in}
\htedit{Our evaluation of the traceback system starts from testing it against
  dirty-label poisoning attacks. We consider three state-of-the-art
  dirty-label attacks:  BadNet~\cite{gu2017badnets}, Trojan~\cite{liu2018trojaning}, and
  Physical Backdoor~\cite{wenger2021backdoor}.  We follow the original
  papers to implement these attacks and vary their attack
  parameters to produce a rich collection of poisoning attacks and
  their misclassification events.   Overall, our results show that
  the proposed traceback system can accurately identify the root cause
  of dirty-label attacks ($\geq 98.9\%$ precision and $\geq 97.1\%$ 
recall) while maintaining a reasonable
traceback time per attack.}

\begin{table}[t]
  \centering
  \resizebox{0.5\textwidth}{!}{
  \begin{tabular}{lcccc}
  \hline
  \textbf{Dataset} & \textbf{Dimensionality} & \textbf{\# Classes} & \textbf{\# Training Data} & \textbf{Architecture} \\ \hline
  CIFAR10~\cite{krizhevsky2009cifar} & $32 \times 32$ & $10$ & $50,000$ & WideResNet-28~\cite{zagoruyko2016wide} \\
  ImageNet~\cite{deng2009imagenet} & $299 \times 299$ & $1,000$ & $1,281,167$ & Inception ResNet~\cite{szegedy2017inception} \\
  VGGFace~\cite{parkhi2015deep} & $224 \times 224$ & $2,622$ & $2,622,000$ & VGG-16~\cite{simonyan2014very} \\
  Wenger Face~\cite{wenger2021backdoor} & $224 \times 224$ & $10$ & $762$ & ResNet-50~\cite{he2016deep} \\
  EMBER Malware~\cite{anderson2018ember} & $2351$ & $2$ & $600,000$ & EmberNN~\cite{severi2021explanation} \\ \hline
  \end{tabular}
  }
  \caption{Datasets \& DNN architectures for our evaluation.}
  \label{tab:task_detail}
\end{table}

\begin{table*}
\centering
\resizebox{0.7\textwidth}{!}{
\begin{tabular}{cccccc}
\hline
\multirow{2}{*}{\textbf{Attack Type}} & \multirow{2}{*}{\textbf{Attack Name}} & \multirow{2}{*}{\textbf{Dataset}} & \multicolumn{3}{c}{\textbf{Traceback Performance}} \\ \cline{4-6} 
 &  &  & \textbf{Precision} & \textbf{Recall} & \textbf{Runtime (mins)} \\ \hline

\multirow{4}{*}{\begin{tabular}[c]{@{}c@{}}Dirty-label\\ (\S\ref{sec:eval})\end{tabular}} & BadNet &  CIFAR10 & $99.5 \pm 0.0\%$ & $98.9 \pm 0.0\%$ & $11.2 \pm 0.4$ \\
& BadNet &   ImageNet& $99.1 \pm 0.0\%$ & $99.1 \pm 0.0\%$ & $142.5 \pm 4.1$ \\
& Trojan &  VGGFace & $99.8 \pm 0.0\%$ & $99.9 \pm 0.0\%$ & $208.9 \pm 9.2$ \\
& Physical Backdoor & Wenger Face & $99.5 \pm 0.1\%$ & $97.1 \pm 0.2\%$ & $2.1 \pm 0.0$ \\ \hline

\multirow{5}{*}{\begin{tabular}[c]{@{}c@{}}Clean-label\\ (\S\ref{sec:clean-label})\end{tabular}} & BP & CIFAR10 & $98.4 \pm 0.1\%$ & $96.8 \pm 0.2\%$ & $19.2 \pm 1.2$ \\
& BP & ImageNet & $99.3 \pm 0.0\%$ & $97.4 \pm 0.1\%$ & $202.0 \pm 7.1$ \\
& WitchBrew & CIFAR10 & $99.7 \pm 0.0\%$ & $96.8 \pm 0.1\%$ & $21.4 \pm 2.1$ \\
& WitchBrew & ImageNet & $99.1 \pm 0.1\%$ & $97.9 \pm 0.1\%$ & $194.3 \pm 5.9$ \\
& Malware Backdoor & Ember Malware & $99.2 \pm 0.0\%$ & $98.2 \pm 0.1\%$ & $57.7 \pm 3.0$ \\ \hline

\end{tabular}
} \vspace{-0.1in}
\caption{Precision, recall, and runtime of the traceback system for each of the four \textbf{dirty-label poisoning} attack tasks and the five \textbf{clean-label poisoning} attack tasks (averaged over $1000$ runs per attack task). }
\vspace{-0.05in}
\label{tab:eval_results}
\end{table*}

\secspace
\subsection{Experiment Setup}
\vspace{-0.06in}
\label{subsec:setup-dirty}
We first summarize  the configuration of the attacks and our
traceback system, and discuss the evaluation metrics.

\para{Attack Setup.} Using 4 image classification datasets listed in Table~\ref{tab:task_detail}, we implement the above mentioned three
attacks (Table~\ref{tab:attack_detail} in Appendix). Their attack success rate and normal
classification accuracy match those reported by the original
papers. We briefly summarize them below. Further details on the DNN models, datasets, and attacks can be
found in Appendix \ref{appsec: exp_details}.


\begin{packed_itemize} \vspace{-0.05in}

\item \textbf{BadNet (CIFAR10,  ImageNet): } 
  The BadNet~\cite{gu2017badnets} attack builds poison training data
  by adding a pre-selected backdoor trigger to benign inputs and
  labeling them with the target label. We run BadNets on image
  classification tasks trained on CIFAR10 and ImageNet, respectively.
  The default attack configuration is identical to~\cite{gu2017badnets}: $10\%$ injection rate (\ie 10\% of the
  training data is poison data) and a `yellow square' as the 
  trigger.


\item \textbf{Trojan (VGGFace): } The Trojan~\cite{liu2018trojaning} attack improves upon 
BadNet by using an optimized trigger to increase 
attack success. Like the original paper~\cite{liu2018trojaning}, we
implement this attack on a face recognition model trained on
VGGFace. The default attack configuration uses $10\%$ injection rate and 
a $59 \times 59$ pixel trigger. 


\item \textbf{Physical Backdoor (WengerFace): } Wenger 
  \etal~\cite{wenger2021backdoor} recently proposed a physical
  backdoor attack against facial recognition models, using everyday
  physical objects such as eyeglasses and headbands as the
  trigger. They collected a custom dataset of face images (hereby referred to
  as WengerFace) of users wearing these accessories.  We use the same
  dataset\footnote{
    We contacted the authors of~\cite{wenger2021backdoor} to obtain
    the dataset and the required 
    user consent and authorization to use this dataset for our study.} to implement the attack. Following the original
paper, we implement the attack with the default configuration of
$10\%$ injection rate and a pair of eyeglasses as the trigger (since
it offers the highest success rate among the triggers tested).
Note that this backdoor attack is able to bypass $4$ state-of-the-art 
defenses~\cite{wang2019neural,tran2018spectral,chen2018detecting,
gao2019strip}. To the best of our knowledge, there is no known
effective 
defense against this attack. 
\vspace{-0.1in}
\end{packed_itemize}
In the rest of the paper, we often use {attack-dataset} (\eg \textsf{
  BadNet-CIFAR10}) to succinctly identify each attack task.
Given an attack configuration (\ie attack-dataset, injection rate,
trigger),  we generate $1000$ successful attack instances $(x_a,y_a)$ as the misclassification events to test 
our traceback system.  Specifically, we randomly choose $10$ target
labels to implement 10 versions of the given poisoning attack. Then for each
attack version, we randomly select $100$ successful attack
instances  as the misclassification events, producing a total of 
$10\times100=1000$ events.

\para{Traceback Setup.} Configuring the traceback system is simple as
it does not assume prior knowledge of the attacks.  The majority of
computation comes from the data projection used by the
clustering component, \ie computing eq. (\ref{eq:reverse-gradient})
for each training sample.  For large models (like those trained on ImageNet), we
speed up the computation of 
Eq. (\ref{eq:reverse-gradient}) by randomly selecting $\rho$\% of the model
weights in the final classification layer.  \htedit{We empirically find that
reducing $\rho$ from $10$ to $1$ does not lead to visible changes
to the traceback accuracy (see Table~\ref{tab:p_kept} in the
Appendix), and thus set $\rho=1$.}


\para{Evaluation Metrics.} We evaluate the proposed traceback system using three 
metrics: 1) \emph{precision} of identifying poison training data, 2) 
\emph{recall} of identifying poison training data, and 3) runtime
latency 
of a successful traceback. We report the average and standard 
deviation values  over $1000$ misclassification events
per attack configuration.

\secspace
\subsection{Traceback Performance}
\vspace{-0.1in}
\label{subsec: traceback_perf_dirty}
\para{Precision and Recall.} We first report the precision and recall of our traceback
system, when tested against dirty-label attacks under the default attack
configuration (\ie those used by the original papers).  The top
section of Table~\ref{tab:eval_results} shows the traceback precision
and recall for each attack-dataset. Across all these experiments, our
system consistently achieves a high precision~($99.1-99.8\%$)
and a high recall~($97.1-99.9\%$). 


An interesting observation is that the recall for Physical Backdoor is lower ($97.1$\%) than the other attacks ($>98.9$\%). That is,
our traceback detects less portion of the 
poison training data samples used by Physical Backdoor compared to
the other attacks.  We wonder whether this is because those data samples
contributed very little to the injected vulnerability, especially since
real photos of physical objects often lead to less precise triggers than those
injected digitally.  We validate this hypothesis by removing the exact
set of poison data missed by our traceback and re-launching the poison
attack, and the attack success rate drops by only 0.08\% on average.
But when removing a random set of poison training
data of the same size,  the attack success rate drops by
1.02\% on average. This confirms our hypothesis. 



\begin{figure*}[t]
  \centering
  \subfigure[\textsf{BadNet-CIFAR10}]{
    \includegraphics[width=0.3\textwidth]{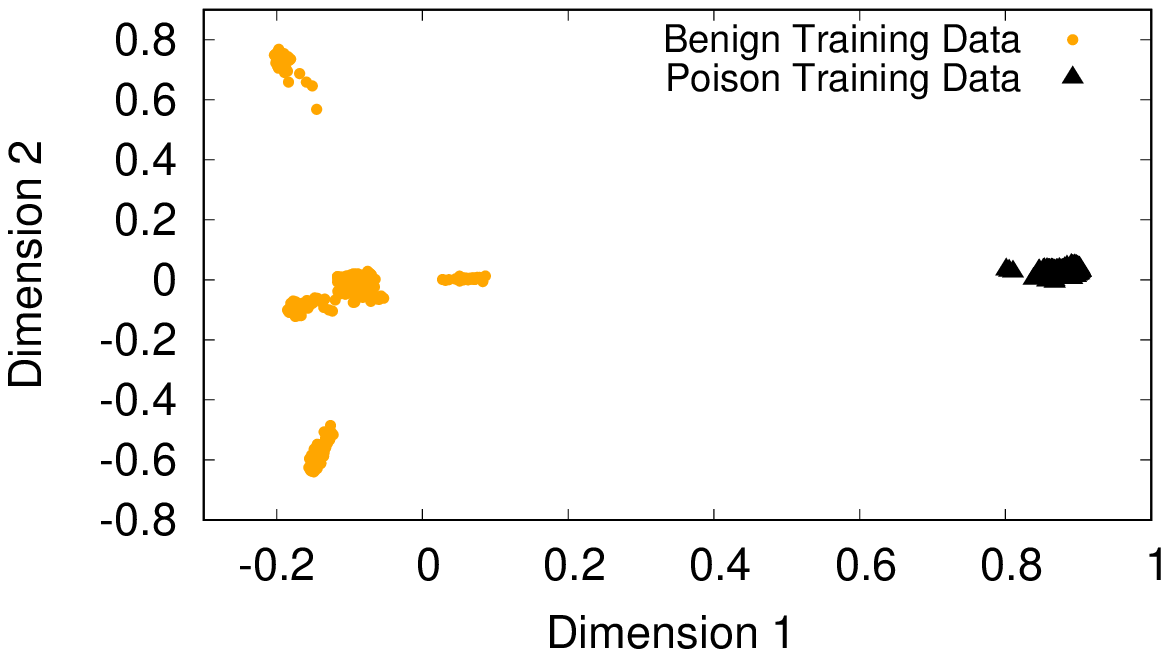}
  }
  \subfigure[\textsf{Trojan-VGGFace}]{
    \includegraphics[width=0.3\textwidth]{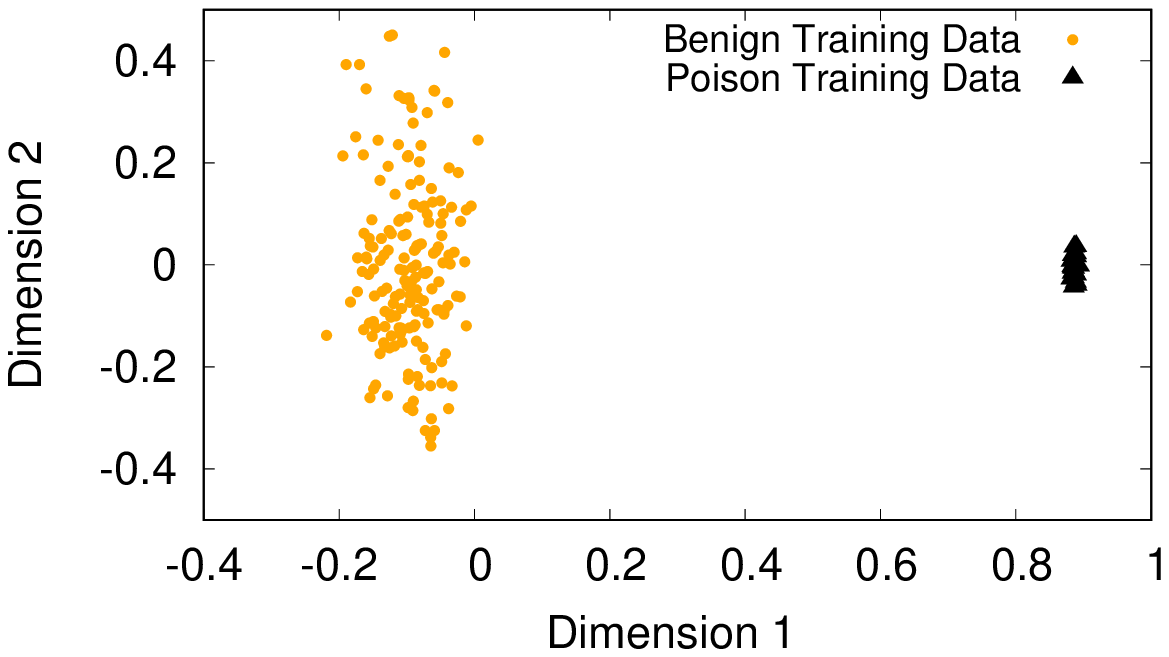}
  }
  \subfigure[\textsf{Physical-Wenger}]{
    \includegraphics[width=0.3\textwidth]{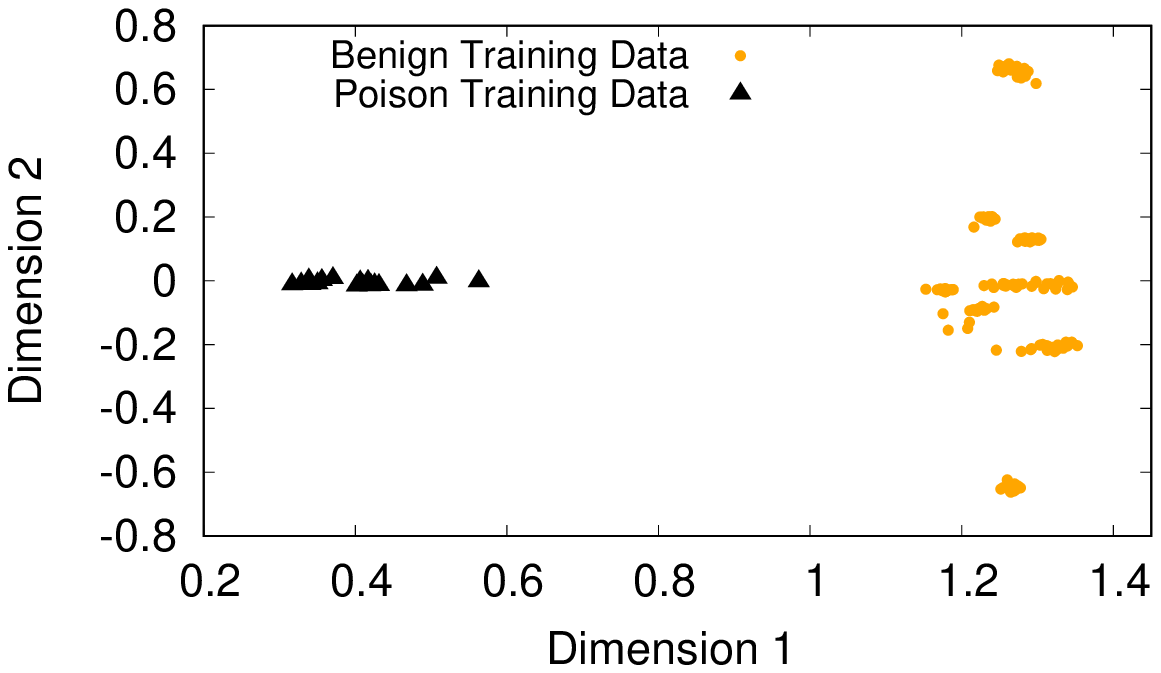}
  }
  \vspace{-0.1in}
  \caption{A simplified, 2-D PCA visualization of the projected training
    data, where poison and benign data are well-separated. }
  \label{fig:pca_dirty}
  \vspace{-0.1in}
\end{figure*}

\para{Detailed Analysis of Clustering.} To obtain a deeper
understanding of 
the traceback performance,  we perform a detailed analysis of the
clustering component. Intuitively, clustering is most effective if the
data projection makes the benign and poison data well-separated from
each other.  Along this line, our analysis starts from visualizing the
project result of the model training data (benign and poison), for three poisoning
attacks (\textsf{BadNet-CIFAR10}, \textsf{Trojan-VGGFace}, 
and \textsf{Physical-Wenger}).  Figure~\ref{fig:pca_dirty} shows the
simplified 2D version generated using 2-dimensional Principal 
Component Analysis (PCA)~\cite{pearson1901liii} on the projected
data.  In this visualization, the benign and poison
data appear to be well-separated.



Next,  we study the $L_2$ distance among the projected training
data, focusing on measuring the normalized $L_2$ distance between each poison
data sample to the centroid of all the benign data (\ie the 
benign centroid), and the centroid of all the poison data (\ie the poison
centroid).  Table~\ref{tab:dist_cluster} lists the average results for four
attacks. We calculate the normalized $L_2$ distances to allow 
a fair comparison across attacks.

Results in Table~\ref{tab:dist_cluster} confirm that the poison and
benign data are reasonably separated.  The poison data in \textsf{Physical-Wenger} are 
more spread out while those in
\textsf{Trojan-VGGFace} are densely packed. These observations align
with those of Figure~\ref{fig:pca_dirty}). Overall, our proposed data
projection achieves sufficient separation between the benign and
poison data, allowing the subsequent clustering and pruning operation
to quickly identify all the benign data. Across all four attacks, the
traceback takes no more than $4$ clustering/pruning iterations to
complete.




\para{Detailed Analysis of Pruning.}  Our pruning operation is based on the condition defined by eq. (\ref{eq:binarymetric}) that compares
the cross-entropy loss of the misclassification event on the original
model $\mathcal{F}$
and the new model $\mathcal{F}^-$ after removing a cluster
$D_1$ from the training dataset. Using \textsf{BadNet-CIFAR10} as an example, 
Table~\ref{tab:pruningcrossentropy} lists the mean and standard deviation of
$\ell (\mathcal{F}(x_a),y_a)$ for the original model, $\ell (\mathcal{F}^-(x_a), y_a)$ when removing
an ``innocent'' cluster, 
and $\ell (\mathcal{F}^-(x_a), y_a)$ when removing a poison
cluster.  The latter two display distinct difference when compared to the first term ($\ell
(\mathcal{F}(x_a),y_a)$), confirming that our proposed binary condition offers a clear
signal to accurately identify innocent clusters not responsible for
the misclassification. 

\begin{table}[h]
\centering
\resizebox{0.4\textwidth}{!}{
\begin{tabular}{l|cc}
\hline
\multirow{2}{*}{$\ell (\mathcal{F}(x_a),y_a)$} &
                                                 \multicolumn{2}{c}{$\ell
                                                 (\mathcal{F}^-(x_a),y_a)$
                                                 \textbf{when removing}} \\ \cline{2-3}
 & \textbf{an innocent cluster} & \textbf{a poison cluster}   \\ \hline
$0.09 \pm 0.02$ &  $0.02 \pm 0.00$ & $6.91 \pm 0.6$ \\ \hline 
\end{tabular}
}
\vspace{-0.05in}
\caption{The cross-entropy loss of the misclassification event on the
  original and modified models, for \textsf{BadNet-CIFAR10}. }
  \vspace{-0.15in}
\label{tab:pruningcrossentropy}
\end{table}



\secspace
\subsection{Traceback Overhead}
\vspace{-0.05in}
\label{subsec:overhead-dirty}
Finally, we report in Table~\ref{tab:eval_results} (the last column) the runtime of traceback against different attack
tasks. We run a prototype of our traceback system on a machine 
with one Nvidia Titan X GPU and $12$ Intel Xeon CPUs. The 
computation time linearly increases with the dimension of the data
projection and the number of training data samples. For models with a large training
dataset, the bulk of the traceback computation comes from the clustering of training 
data, which takes up $83\%$ of the computation time for
\textsf{Trojan-VGGFace}, the most computational expensive task. On the
other hand, simple data 
parallelism enabled by the use of mini-batch 
K-mean clustering can significantly speed up the runtime. For example,
parallelizing using $5$ machines reduces the runtime for \textsf{Trojan-VGGFace} to $49.5 \pm 3.2$ 
minutes, a $4.1$x speed up.

\secspace
\section{Evaluation on Clean-label Attacks}
\label{sec:clean-label}
\vspace{-0.06in}
We now evaluate our traceback system against 
clean-label poisoning attacks, and contrast its performance to that 
on dirty-label attacks. Compared to dirty-label attacks, clean-label
attacks follow a different attack methodology,  use fewer poison
training samples, and these samples appear
less separated from the benign data even after data projection. These
factors make the clustering and pruning process more
challenging.  Nevertheless, our traceback system still achieves good
performance across all attack tasks ($\geq 98.4\%$ precision and $\geq
96.8\%$ recall).  In the following, we present the experiment setup of 
the five clean-label attacks used by our evaluation, and how our traceback system responds to
these attacks.


\secspace
\subsection{Experiment Setup} 
\vspace{-0.05in}

\para{Attack Setup.} We consider two state-of-the-art 
triggerless clean-label attacks on image classification, and 
a backdoor-based clean-label attack on malware classification 
(Table~\ref{tab:clean_label_detail} in Appendix). Our implementation of these 
attacks match the original papers in both attack success rate and benign classification 
rate. Further implementation details are in 
Appendix~\ref{appsec: exp_details}.



\begin{packed_itemize}\vspace{-0.05in}

\item \textbf{Bullseye Polytope (BP) (CIFAR10 \& ImageNet):}  The BP
  attack~\cite{aghakhani2020bullseye} aims to make the model classify 
a single attack sample $x_a$ to a target class $y_a$ at test time without 
modifying the sample (hence, triggerless). This is done by adding  
imperceptible perturbations to the poison training data so their 
representations in the feature space form a fixed-radius polytope 
around the chosen attack sample $x_a$. The classifier then associates the 
region around the attack sample with the label of the poison data, 
which is the desired target label $y_a$. This leads to 
misclassification of $x_a$ to $y_a$ at test time. 
  
It is known that BP only works well when the attacker has access to a 
pretrained feature extractor used by the victim model, \ie it is effective 
in the transfer learning setting. We follow this setup and use a 
feature extractor pretrained on the benign data as the victim. We 
use the attack parameters of the BP-5x attack (the strongest 
variant) from the original paper and test the attack on  CIFAR10 and 
ImageNet.
  


\item \textbf{Witches' Brew (CIFAR10 \& ImageNet):} Witches' 
Brew~\cite{geiping2020witches} is a clean-label triggerless attack 
that. It works by adding imperceptible perturbations to the 
poison data to align its gradient with that of the attack sample. 
This makes the model misclassify the attack sample to the 
target class of the poison data. We test the attack on 
CIFAR10 and ImageNet datasets. 


\item \textbf{Malware Backdoor (Ember Malware):} This is a 
clean-label, \emph{backdoor} attack on malware classifiers~\cite{
severi2021explanation}. The attacker uses influence 
functions to find $128$ most important features defining `goodware' and 
uses these as a trigger. These features are then modified for 
poison malware samples, with the target class being `goodware'. 
The authors of~\cite{
severi2021explanation}  found all three of the state-of-the-art defenses~\cite{
chen2018detecting,liu2008isolation,tran2018spectral} are ineffective 
against this attack. Like the original paper,  we use the 
publicly-available Ember malware classification dataset. Since 
attackers are mostly interested in disguising malware as `goodware', 
we only use `goodware' as the target label of the attack.
  \vspace{-0.1in}
\end{packed_itemize}

\para{Traceback System Setup \& Evaluation Metrics. } We use the same 
system setup and evaluation metrics as \S\ref{sec:eval}.


\secspace
\subsection{Traceback Performance}
\vspace{-0.05in}

The bottom section of Table~\ref{tab:eval_results} 
shows that our traceback system achieves $>98.4\%$ precision and 
$>96.8\%$ recall when going against the five clean-label poisoning
attacks.  In the following, we discuss these results in detail by contrasting the trackback performance on 
clean-label attacks to that on dirty-label attacks 
(discussed in \S~\ref{subsec: traceback_perf_dirty}).

\para{Lower traceback recall due to ineffective poison data.} First,  we see that the traceback precision remains high even as compared to dirty-label 
attacks, but the recall is consistently lower and the runtime is higher.  In
particular, Table~\ref{tab:eval_results}  shows that the lowest
traceback recall happens on the two
triggerless attacks, BP and Witches' Brew.  We hypothesize that it is 
because these two attacks failed to move the 
representations of some poison training data to the desired location in the 
feature space. These ``ineffective'' poison training data made very little
contribution to the
misclassification event, and are hard to detect during traceback. 


We test and validate this hypothesis by 
gradually increasing the attack perturbation budget (to increase the
effectiveness of the poison training data) and observing the
attack success rate. 
The budget for the BP and WitchBrew attacks determines the magnitude 
of perturbation the attacker can add to each poison data point (the
default budget is set to $L_{inf} = 0.03$). A higher perturbation 
budget allows the attacker to position the poison training data ``closer'' to their 
desired locations, making these data more ``effective'' and leading to a stronger attack success rate. 
Table~\ref{tab:budget_bp} confirms that the attack success rate increases 
with the perturbation budget.  We also list the traceback precision and
recall.  While precision remains high, the 
recall also increases from $94.9\%$ to $99.4\%$ as we increase the
budget, confirming our hypothesis that the presence of ineffective
poison samples is 
the reason for a lower recall value. 


\para{Less distinct clusters lead to more pruning iterations.} We also
study the performance of clustering and pruning on the five clean-label
attacks. Like
Figure~\ref{fig:pca_dirty} for dirty-label attacks, we visualize the
projected training data using 2-D PAC for clean-label attacks.  The
visualization for \textsf{BP-CIFAR10} is shown in
Figure~\ref{fig:cluster_bp_pca}.  Compared to the wide separation seen
on the dirty-label poison data (Figure~\ref{fig:pca_dirty}), the BP
poison data appear much less separated from the benign training
data. In fact, they reside in the gap between different benign clusters.
This is not surprising, because clean-label attacks work by moving the
representation of the poison data close to that of benign data in the
target class,  so these poison data are inherently closer to the
benign data.


  We also analyze the pruning operation when tracing back clean-label
  attacks. Table~\ref{tab:pruningcrossentropy-bp} shows the
  cross-entropy loss caused by training data removal, which is used to determine whether a cluster is benign or
  not.  Again, we observe a clear pattern that separates benign
  clusters from poison (or mixed) clusters, \ie $\ell
  (\mathcal{F}^-(x_a), y_a)$ for removing an innocent cluster is
  smaller than $\ell (\mathcal{F}(x_a)$, and $\ell
  (\mathcal{F}^-(x_a), y_a)$ for removing a poison cluster is
  significantly larger.  As such, pruning can effectively identify
  benign clusters. 




\begin{table}[t]
\centering
\resizebox{0.4\textwidth}{!}{
\begin{tabular}{l|cc}
\hline
\multirow{2}{*}{$\ell (\mathcal{F}(x_a),y_a)$} &
                                                 \multicolumn{2}{c}{$\ell
                                                 (\mathcal{F}^-(x_a),y_a)$
                                                 \textbf{when removing}} \\ \cline{2-3}
 & \textbf{an innocent cluster} & \textbf{a poison cluster}   \\ \hline
$0.61 \pm 0.07$ &  $0.39 \pm 0.04$ & $8.81 \pm 0.81$ \\ \hline
\end{tabular}
}
\caption{The cross-entropy loss of the misclassification event on the
  original and modified models, for \textsf{BP-CIFAR10}. }
\label{tab:pruningcrossentropy-bp}
\end{table}

Despite the reduced separation,  our traceback system can still
accurately identify the benign (and thus poison) data clusters while
using more 
clustering/pruning iterations.  For example, an average of $10.7$ iterations are needed for \textsf{BP-CIFAR10} 
compared to $3.2$ iterations for the dirty-label attack on the same
model (\textsf{BadNet-CIFAR10}).

\begin{table}[t]
\centering

\vspace{-0.06in}
\resizebox{0.48\textwidth}{!}{
\begin{tabular}{lccc}
\hline
\multirow{2}{*}{\textbf{Attack-Dataset}} & \multicolumn{3}{c}{\textbf{Traceback Method}} \\ \cline{2-4} 
 & Spectral Signature & Neural Cleanse & \textbf{Ours} \\ \hline
BadNet-CIFAR10 & 95.3\% / 92.4\% & 98.7\% / 96.6\% & \textbf{99.5\% / 98.9\%} \\
BadNet-ImageNet & 96.0\% / 93.7\% & 91.1\% / 97.2\% & \textbf{99.1\% / 99.1\%} \\
Trojan-VGGFace & 93.1\% / 89.8\% & 94.8\% / 97.4\% & \textbf{99.8\% / 99.9\%} \\
Physical-Wenger & 43.2\% / 67.4\% & 0\% / 0\% & \textbf{99.5\% / 97.1\%} \\
Malware Backdoor & 2.1\% / 15.1\% & 0\% / 0\% & \textbf{99.2\% / 98.2\%} \\ \hline
\end{tabular}

}
\vspace{-0.05in}
\caption{\shawn{Comparing our traceback system against forensic tools adapted
    from existing backdoor defenses. We present the results as ``Precision / Recall''.}}
\label{tab:comparsion_backdoor}
\vspace{-0.1in}

\end{table}

\begin{table}[t]
\centering

\vspace{-0.06in}
\resizebox{0.48\textwidth}{!}{
\begin{tabular}{lccc}
\hline
\multirow{2}{*}{\textbf{Attack-Dataset}} & \multicolumn{3}{c}{\textbf{Traceback Method}} \\ \cline{2-4} 
 & Deep K-NN & $L_2$-Norm & \textbf{Ours} \\ \hline
BP-CIFAR10 & 36.1\% / 74.3\% & 34.5\% / 78.0\% & \textbf{98.4\% / 96.8\%} \\
BP-ImageNet & 57.9\% / 79.6\% & 53.4\% / 72.4\% & \textbf{99.3\% / 97.4\%} \\
WitchBrew-CIFAR10 & 49.3\% / 53.9\% & 52.1\% / 42.8\% & \textbf{99.7\% / 96.8\%} \\
WitchBrew-ImageNet & 53.5\% / 47.2\% & 51.3\% / 44.3\% & \textbf{99.1\% / 97.9\%} \\ \hline
\end{tabular}

}
\caption{\shawnmr{Comparing our traceback system against forensic tools adapted from existing clean-label defenses.
We present the results as ``Precision / Recall''.}}

\label{tab:comparsion_clean}
\vspace{-0.1in}

\end{table}

\shawnmr{
\secspace
\section{Comparing against Adapted Defenses}
\vspace{-0.05in}
\label{subsec:compare}

While forensics tools are solving a different problem as defenses, it is
reasonable to ask, can existing defenses be adapted to become tools in
forensic analysis. Here, we adapt four state-of-the-art poison defenses (two
backdoor and two triggerless defenses) to perform post-attack traceback
analysis. We compare our system against adapted backdoor defenses (Spectral
Signature~\cite{tran2018spectral} and Neural Cleanse~\cite{wang2019neural}) in
Table~\ref{tab:comparsion_backdoor}, and our system against triggerless
defenses (Deep K-NN and $L_2$-Norm Outliers~\cite{peri2020deep}) in
Table~\ref{tab:comparsion_clean}. 

\para{Spectral Signature.} Spectral signature~\cite{tran2018spectral} extracts signatures
for backdoor training data using the spectrum of the covariance of
the feature representation. Spectral signature identifies the 
malicious training samples post-training (before attack). We adapt 
spectral signature simply by using it to identify the
malicious training data post-attack.

Table~\ref{tab:comparsion_backdoor} shows that spectral signature performs
well on BadNet and Trojan attacks, but performs quite poorly when tracing
back attacks for physical backdoors and malware backdoors. The poor
performance against Malware Backdoors is consistent with the malware paper
itself~\cite{severi2021explanation}, where the authors showed that spectral 
signature defense is ineffective against their proposed malware attack.

\para{Neural Cleanse (NC). } NC~\cite{wang2019neural} recovers backdoor
triggers in a poisoned model by reverse-engineering the backdoor trigger.
For traceback, we apply NC and search for the recovered trigger in the
training dataset to identify poison data. Since NC does not recover the exact
input trigger, we perform the matching in neuron activation space, \ie
flagging the training data if its neuron activation is close to that of the
recovered trigger. We use cosine similarity to measure the activation
distance and use a small set of benign data to calculate a cutoff threshold.

In our tests, NC performs well on BadNet and Trojan backdoors, where it successfully 
recovers backdoor triggers. Against physical and malware backdoor attack, however,
NC fails to identify any triggers, and thus traceback fails.

\para{Deep K-NN and $L_2$-Norm Outliers. } Both defenses are proposed
by~\cite{peri2020deep}, where clean-label poison data are detected using
anomaly detection in the feature space.  In our tests, both defenses perform
poorly on all four clean-label attack tasks
(Table~\ref{tab:comparsion_clean}), achieving similar performance as random
guessing.  This result is also consistent with results from the BP and
WitchBrew attack papers, where they show existing defenses are not
effective~\cite{aghakhani2020bullseye,geiping2020witches}.}

\begin{figure*}
  \centering
  \includegraphics[width=0.97\textwidth]{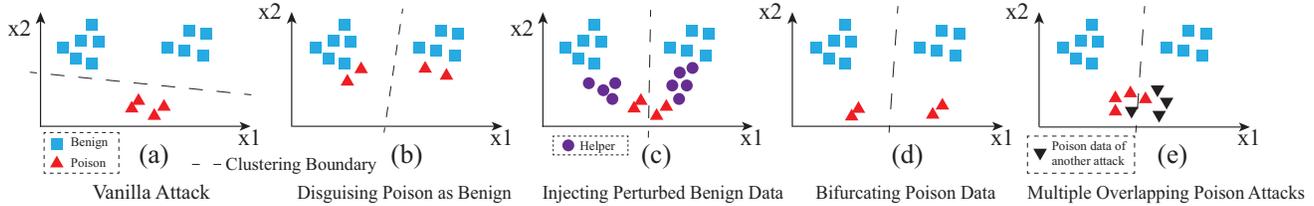}
  \vspace{-0.1in}
  \caption{An illustration of four countermeasures where attacker can
    manipulate the data layout in data projection space in order to disrupt traceback.} 
  \label{fig:counter_illu}
  \vspace{-0.1in}
\end{figure*}

\secspace
\section{Anti-Forensic Countermeasures}
\label{sec:counter}

As with other security mechanisms, we expect attackers with knowledge of our
system to deploy adaptive countermeasures. In this section, we explore
potential anti-forensics techniques and evaluate them for their impact on our
system. 


Our experiments make the strongest possible assumption about the attacker:
that they know full details of the target model's training pipeline,
including read access to all training data, model architecture, and training
hyperparameters. They also know details of the traceback system, including
the function to calculate the data projection. We assume that attacker can add
additional perturbations to its poison data in order to evade traceback, and
we assume a generous perturbation budget of $L_{inf} = 0.1$ (attacker can
change each input value by $10\%$). For trigger-based attacks, we further
allow the attacker to arbitrarily change the location and value of the
triggers.

We consider \shawnmr{five} different countermeasures, each leveraging in different ways
the attacker's ability to inject manipulated training data into the data
pipeline. The countermeasures are shown in Figure~\ref{fig:counter_illu}, and
include: a) disguising poison data as benign, 
b) injecting perturbed benign data, c) bifurcating poison data, and d)
multiple overlapping poison attacks.  We evaluate the countermeasures on a
total of $7$ attacks, all of our attacks except for physical backdoors (cannot easily
modify physical triggers) and malware backdoors (fixed triggers based on
heuristic). \shawnmr{We also test the scenarios where attacker increases  
the attack's cost in order to evade traceback. }



\para{Disguising Poison as Benign.} 
\label{subsec:disguise}
The first countermeasure tries to confuse the clustering algorithm, by
perturbing the poison data and pushing it closer to benign data in the data
projection space (Figure~\ref{fig:counter_illu}b).  If the clustering
heuristic groups poison data into different benign clusters, it will
terminate the pruning and either reduce recall or confuse the system into
identifying it as a non-poison misclassification.

We test the case where an attacker minimizes the
distance between each poison data and a benign data point closest to it in
the data projection space, while optimizing for the original attack objectives, 
using a $\lambda$ parameter to balance the two objectives. The attacker
leverages a \emph{bi-level} optimization~\cite{huang2020metapoison} to
optimize the attack objective, since the modifications of poison data also
change the result model and thus, the projection calculation.

Results are consistent across all $7$ attack tasks. 
For brevity, we show detailed results for
\textsf{Trojan-VGGFace} (Table~\ref{tab:badnets_disguise}) and
\textsf{BP-CIFAR10} (Table~\ref{tab:bp_disguise}). As the $L_2$ distances
between poison and benign data decrease, it becomes harder for the model to learn the
attack, and attack success rate drops to zero.  Impact on
traceback is minimal: precision and recall drop less than $3.6\%$ across
all tasks. 

\begin{table}[t]
\centering
\resizebox{0.4\textwidth}{!}{
\begin{tabular}{ccccc}
\hline
\textbf{$L_2$ Distance} & \textbf{\begin{tabular}[c]{@{}c@{}}Attack \\ Success Rate\end{tabular}} & \textbf{Precision} & \textbf{Recall} \\ \hline
$434.5\pm 8.2$ & $99.5 \pm 0.0\%$ & $99.5\pm 0.0\%$ & $98.9\pm 0.0\%$ \\
$290.4 \pm 8.9$ & $89.5\pm 0.6\%$ & $98.4\pm 0.1\%$ & $98.1\pm 0.1\%$ \\
$184.8 \pm 5.7$ & $64.3\pm 1.7\%$ & $96.9\pm 0.3\%$ & $97.2\pm 0.1\%$ \\
$110.3 \pm 3.1$ & $28.3\pm 3.2\%$ & $95.9\pm 0.3\%$ & $96.7\pm 0.2\%$ \\
$59.0 \pm 1.9$ & $0.0 \pm 0.0\%$ & N/A & N/A \\ 
$31.4 \pm 1.2$ & $0.0 \pm 0.0\%$ & N/A & N/A \\ \hline
\end{tabular}
}
\caption{For disguising \textsf{Trojan-VGGFace},  attack success rate drops
  as the $L_2$ distance between poison and benign projections decreases,
  while traceback precision and recall drop slightly.} 
\label{tab:badnets_disguise}
\end{table}

\begin{table}[t]
\centering
\resizebox{0.4\textwidth}{!}{
\begin{tabular}{ccccc}
\hline
\textbf{$L_2$ Distance} & \textbf{\begin{tabular}[c]{@{}c@{}}Attack \\ Success Rate\end{tabular}} & \textbf{Precision} & \textbf{Recall} \\ \hline
$28.2\pm 0.9$ & $86.1 \pm 1.4\%$ & $98.4 \pm 0.1\%$ & $96.8 \pm 0.2\%$ \\
$25.4 \pm 1.3$ & $59.4 \pm 3.4\%$ & $97.2\pm 0.2\%$ & $95.0\pm 0.3\%$ \\
$19.0\pm 1.2$ & $19.7 \pm 2.0\%$ & $96.1\pm 0.2\%$ & $94.8\pm 0.3\%$ \\
$12.8\pm 0.8$ & $8.7 \pm 0.4\%$ & $95.9\pm 0.1\%$ & $95.1\pm 0.3\%$ \\
$9.4\pm 0.5$ & $0.0 \pm 0.0\%$ & N/A & N/A \\ 
$4.4\pm 0.3$ & $0.0 \pm 0.0\%$ & N/A & N/A \\ \hline
\end{tabular}
}
\caption{For disguising \textsf{BP-CIFAR10}, attack success rate decreases as
  $L_2$ distance between poison and benign projections decreases, while
  traceback precision and recall drop slightly.}
\label{tab:bp_disguise}
\end{table}

\shawnmr{

  \para{Targeted Contamination Attack (TaCT). } Another method to disguise
  poison data is the TaCT attack~\cite{tang2021demon}. TaCT is designed to
  obscures the difference between poison and benign representations in order
  to evade existing defenses. TaCT leverages source-specific backdoors,
  triggers \textit{only} work when applied to specific images.  The authors
  found that TaCT causes poison and benign data to be inseparable in neuron
  representation space.

  We test our traceback system against TaCT. We find that while
  poison and benign data are inseparable in neuron representation space, we
  can still separate them in our data projection space. We believe the
  separation is likely because our data projections measure each data's
  \textit{impact} on model parameters, which better characterize training data
  than neurons activation values in poison settings. We test TaCT attack
  on all $4$ backdoor attack tasks from \S\ref{sec:eval}. Overall, TaCT
  attack is ineffective against our traceback system, reducing precision
  and recall by less than $4\%$ across the five backdoor attacks.  }

\para{Injecting Perturbed Benign Data.} Next, we consider the countermeasure where the attacker seeks to confuse the
clustering algorithm by injecting benign data whose projection fills in the
space between benign and poison data.  We refer to
this additional set of modified data as ``helper data''. We note that the
helper data do not directly contribute to the success of the
misclassification event.  

As shown in Figure~\ref{fig:counter_illu}(c), an attacker  positions
helper data between poison and benign data to mislead the
clustering heuristic.  The attacker first identifies the \emph{last} set of benign data
pruned out by the traceback system. Using the clustering algorithm, the
attacker separates the benign cluster into two clusters,  and the poison data
into two groups based on proximity to each benign cluster's centroid.
The attacker optimizes helper data to uniformly position them
in between each benign cluster and its closeby poison cluster. The attacker
uses a similar bi-level optimization (\S\ref{subsec:disguise}) to optimize
the attack objective.

We apply this countermeasure on the $7$ attack tasks. For brevity, we 
show detailed results for \textsf{Trojan-VGGFace} in
Table~\ref{tab:trojan_helper}. As the number of helper data samples increases to
$10000$ ($25\%$ of training data), attack success rate reduces gradually
to zero, while traceback precision drops to $94.1\%$ and recall
remains the high ($> 98.7\%$). Table~\ref{tab:bp_helper} shows 
results for \textsf{BP-CIFAR10}, where the attack success rate drops much
faster when injecting merely $25$ helper data. The faster drop in attack
success is likely due to the fewer clean-label poison samples and
their proximity to benign data. When Trojan and BP attacks reach zero attack
success rate ($15000$ and $20$ helper data respectively), our traceback can
still separate poison data with $> 91.4\%$ precision and recall. 

\begin{table}[t]
\centering
\resizebox{0.40\textwidth}{!}{
\begin{tabular}{rccc}
\hline
\textbf{\begin{tabular}[c]{@{}c@{}}Number of \\ Helper Data\end{tabular}} & \textbf{\begin{tabular}[c]{@{}c@{}}Attack \\ Success Rate\end{tabular}} & \textbf{Precision} & \textbf{Recall} \\ \hline
0 & $99.8 \pm 0.0\%$ & $99.8 \pm 0.0\%$ & $99.9 \pm 0.0\%$ \\
1000 & $64.3 \pm 3.8\%$ & $96.3 \pm 0.2\%$ & $99.1 \pm 0.0\%$ \\
10000 & $21.0 \pm 3.9\%$ & $94.1 \pm 0.3\%$ & $98.7 \pm 0.0\%$ \\ 
15000 & $0.0 \pm 0.0\%$ & N/A & N/A \\\hline
\end{tabular}
}
\caption{For adding helper data to \textsf{Trojan-VGGFace}, the attack success rate decreases as the number of helper data increases, while the precision and recall of the traceback system drop slightly.  }
\label{tab:trojan_helper}
\end{table}

\begin{table}[t]
\centering
\resizebox{0.40\textwidth}{!}{
\begin{tabular}{rccc}
\hline
\textbf{\begin{tabular}[c]{@{}c@{}}Number of \\ Helper Data\end{tabular}} & \textbf{\begin{tabular}[c]{@{}c@{}}Attack \\ Success Rate\end{tabular}} & \textbf{Precision} & \textbf{Recall} \\ \hline
0 & $86.1 \pm 1.4\%$ & $98.4 \pm 0.1\%$ & $96.8 \pm 0.2\%$ \\
5 & $35.5 \pm 3.4\%$ & $96.6 \pm 0.3\%$ & $96.6 \pm 0.1\%$ \\
10 & $13.1 \pm 1.1\%$ & $94.3 \pm 0.5\%$ & $96.6 \pm 0.2\%$ \\
20 & $0.0 \pm 0.0\%$ & N/A & N/A \\ \hline
\end{tabular}
}
\caption{For adding helper data to \textsf{BP-CIFAR10}, the attack success rate decreases as the number of helper data increases, while the recall of the traceback system remains the same and precision drops slightly. }
\label{tab:bp_helper}
\end{table}

\para{Bifurcating Poison Data.} Next, we explore techniques to separate poison data into multiple (two)
separate distributions both of which contribute to the attack incident while
residing in different parts of data projection space, in order to evade clustering
(Figure~\ref{fig:counter_illu}(d)).
The attacker first identifies the two strongest clusters in the poison data, 
then maximize the distance between the cluster centroids to
separate them. We follow the same bi-level optimization process to optimize
the poison data and use a $\lambda$ term to balance the objective of cluster
distance and the original attack objective.  

We apply this countermeasure to all $7$ attack tasks. For BP attacks and
WitchBrew attacks, attack success drops quickly because the two triggerless
attacks rely on clever positioning of the poison data (\eg a fixed radius
polytope around target data). For BP and Witches' Brew, this countermeasure
has no impact on traceback performance ($>97.0\%$ precision and recall). For
trigger-based attacks, we show results on \textsf{Trojan-VGGFace} in
Table~\ref{tab:trojan_separate_two}. We found that as we increase 
$\lambda$ to push for better separation between the two clusters, the
centroid distances fail to increase beyond a certain value. We believe the
failure to separate poison data is because these poison samples
have the same attack objective and trigger, and naturally cluster together in
the data projection space. Overall, the traceback system achieves $\leq 96.7\%$ precision and recall across all $7$ attack tasks. 

\begin{table}[t]
\centering
\resizebox{0.4\textwidth}{!}{
\begin{tabular}{cccc}
\hline
\textbf{$L_2$ Distance} & \textbf{\begin{tabular}[c]{@{}c@{}}Attack \\ Success Rate\end{tabular}} & \textbf{Precision} & \textbf{Recall} \\ \hline
$2.2 \pm 0.2$ & $99.8 \pm 0.0\%$ & $99.8 \pm 0.0\%$ & $99.9 \pm 0.0\%$ \\
$17.9 \pm 2.7$ & $98.3 \pm 0.2\%$ & $98.2 \pm 0.0\%$ & $97.9 \pm 0.1\%$ \\
$25.3 \pm 4.1$ & $97.1 \pm 3.7\%$ & $97.3 \pm 0.2\%$ & $96.9 \pm 0.2\%$ \\
$23.6 \pm 5.8$ & $97.4 \pm 0.0\%$ & $97.5 \pm 0.1\%$ & $97.3 \pm 0.1\%$ \\ 
$24.0 \pm 6.1$ & $98.1 \pm 0.0\%$ & $97.6 \pm 0.2\%$ & $97.0 \pm 0.2\%$\\ \hline
\end{tabular}
}
\caption{For separate one Trojan attack into two, the attack success rate decreases as the $L_2$ distance between centroids decreases, while the precision of the traceback system remains the same and recall drops slightly. }
\label{tab:trojan_separate_two}
\end{table}

\para{Multiple Overlapping Poison Attacks. } 
Finally, an attacker can try to combine two dirty-label attacks in one
misclassification event, by training two different triggers with the same
misclassification label into the model, then 
including both triggers into a single attack input. This attack does not work
for triggerless attacks, since each attack has its own specific target data.  

Our experiments show this countermeasure is \emph{ineffective} against our
traceback system. We achieve $>98.4\%$ precision and recall across all attack tasks. 
While the two poison attacks leverage different
triggers, they have the same objective of misclassifying any inputs to the
same target label, and our data projection directly correlates to the
objective of each training data. Thus poison data from two
separate attacks appear in the same region in projection space, enabling us to
cluster them  together as part of the same attack.

\shawnmr{
\para{Higher Cost Attacks. } So far, we have focused on attacks with a similar 
cost, \ie same number of poison data. Now, we explore the impact of attacks with 
higher cost on our traceback system. We allow an adaptive attacker to poison an 
increasing number of poison data and test our traceback effectiveness against these
higher cost attacks. We found that increasing injection rate has an surprisingly low 
impact on our traceback system. As attacker increases injection rate to $50\%$, 
our traceback system maintains $> 95\%$ precision and $> 92\%$ recall, across 
all $5$ countermeasures discussed in this section and all $9$ attack tasks. 

We believe the weak impact of increasing injection rate is because our traceback system views poison as a group 
effect (\S\ref{sec:intuition}), and poison data with the same attack objective 
are clustered together regardless of the number of poison data. As a result, increasing 
injection rate has limited effectiveness against our traceback system. 
}

\secspace
\section{Discussion: Identifying Non-poison Events}
\label{sec:discussion}


Our work addresses the question of post-attack analysis for poison attacks on
neural networks. In practice, however, a system administrator must first
identify if a misclassification event was caused by a poisoning attack, or
from an \emph{evasion attack} or \emph{benign misclassification}. The
former are test-time attacks that leverage existing vulnerabilities in
trained models to cause misclassification with perturbed data, while the
latter simply arise since models do not classify perfectly.

\para{Attack Identification.} We note that our system can double as a tool
for \shawnmr{the first step towards} attack identification. Given a model and a
misclassification event
(misclassified input and output), one iteration of our forensic system would
be able to identify if the attack was a poison attack or caused by other
means. Once we separate training data into clusters, and apply the same
unlearning techniques, we can observe if removal of either subset of training
data will alter misclassification behavior.

Intuitively, both evasion by adversarial perturbation and benign
misclassification rely on specifics of the model's loss landscape. In either
case, removal or ``unlearning'' of any significant portion of training data
will change the loss landscape and should alter the misclassification
behavior. In our system, we would observe that unlearning either of the
clusters would alter the misclassification event. So if the first iteration
fails to prune away either cluster, then we consider the misclassification
event as non-poison and end traceback.

\minor{

\para{Limitation. } Our attack identification system can be vulnerable 
to ``false flag'' attack~\cite{rid2015attributing} where an attacker carefully crafts a misclassification 
event that triggers our traceback system to blame an innocent data provider. 
This is a threat that the deployer of traceback system needs to keep in mind when 
perform any forms of prosecution based on traceback results. 
In practice, a system like ours must be understood in context: the assignment of blame to any parties involved will only be possible with the availability of an effective data provenance tool, and there should be a human-in-the-loop confirmation before any lasting decisions are made. 

Further, any attacker trying to carry out an inference time attack that also has a false flag component will have to solve a more challenging optimization problem needing access to the training data. The exploration of techniques to do this effectively is beyond the scope of this paper but is an interesting direction for future work. 

}



\para{Empirical Results.} We test four representative \emph{evasion attacks},
including two white-box evasion attacks (PGD~\cite{madry2017towards},
CW~\cite{carlini2017towards}) and two black-box evasion attacks
(Boundary~\cite{brendel2017decision} and
HSJA~\cite{chen2020hopskipjumpattack}) against all $5$ of our evaluation
datasets. We follow the default attack parameters~\cite{shan2020gotta} 
(Appendix~\ref{tab:evasion_details}). We test $100$ evasion attack samples
for each attack and classification task pair.  For \emph{benign
  misclassification}, we randomly select $100$ misclassified benign test data
as the misclassification events for each task. For all these misclassification events, our
forensic tool correctly determined that they were not caused by data
poisoning attacks, \ie our tool produced no false positives.

\secspace
\section*{Future Work}

We believe the study of post-attack forensic techniques remains an open area
with numerous open questions. First, while our method is effective in our
tests, more effort is needed to understand its robustness when extended to
additional types of attacks and application domains. Second, in order to
assign responsibility to a provider for the poisoned data, the model
administrator must be confident that metadata associated with training data
is accurate and tamper-resistant. The study of data
provenance~\cite{fang2011high,lu2010secure} in this context remains an open
problem.

\secspace
\section*{Acknowledgments}
\vspace{-0.05in}
We thank our anonymous reviewers for their insightful
feedback.  This work is supported in part by NSF grants CNS-1949650,
CNS-1923778, CNS-1705042, by C3.ai DTI, and by the DARPA GARD program.  
Shawn Shan is supported in part by a Neubauer Fellowship at the
University of Chicago.  Any opinions, findings, and conclusions or
recommendations expressed in this material are those of the authors and do
not necessarily reflect the views of any funding agencies.

{
 \footnotesize
 \bibliographystyle{plain}
 \bibliography{forensics}
}

\appendix
\section{Appendix}
\label{sec:appendix}


\secspace
\subsection{Theoretical Analysis of Training Data \\ Removal}
\label{appsec: theorem_proof}
As discussed in \S\ref{sec:intuition}, our binary measure of
event responsibility is inspired by a theoretical analysis on how removing a portion of the training data
affects the poisoning attack performance.  We now present this
theoretical analysis in detail. 


To examine the impact of removing a subset of training data ($D_1$) on the poisoning
attack, we seek to analytically quantify its impact on the model
loss over the true distribution of the poison test data,
which indicates the contribution of $D_1$ to successful
misclassification at test time. Furthermore, our analysis considers the general
case where $D_1$ may contain both benign and poison training data, \ie
$D_1$ is drawn from a mixed distribution.  We note that while our
theoretical analysis is driven by the expected value of the loss over the distribution of
the poison test data, in practice the traceback system can only
measure the impact on a single misclassification event (usually just one data sample).  Yet this is
empirically sufficient to label and prune clusters (as shown by our
experimental results in \S\ref{sec:eval} and \ref{sec:clean-label}).  Finally, since our clustering is able to find clusters that only contain benign data, the pruning
component used by our traceback focuses on picking this cluster ($D_1$). This is a special case under our theorems, which show that clusters can be differentiated even if they are mixed.

\para{Definitions:} Let the full training data $D$ be drawn from a
distribution $\mathcal{D}$ comprised of the benign distribution
$\mathcal{D}_b$ and the poison distribution $\mathcal{D}_p$ in the
ratio $\alpha$, \ie $\mathcal{D}= \alpha \mathcal{D}_b+(1- \alpha)
\mathcal{D}_p$.  Let $\mathcal{F}$ denote the original model trained
on $D$, using the loss function $\ell$. To measure the impact of
removing a group of data $D_1$ from the training dataset,  we consider
a new model $\mathcal{F}^-$ trained on $D \setminus D_1$, effectively
drawn from a distribution $\mathcal{D}^-= \alpha^- \mathcal{D}_b+(1-
\alpha^-) \mathcal{D}_p$. Finally, the expected loss over the true distribution for a classifier $\mathcal{F}$ is $L_{\mathcal{D}}(\mathcal{F})$= $\mathbb{E}_{(x,y) \sim \mathcal{D}} \left[  \ell(\mathcal{F}(x),y)  \right]$.



\para{Key Results:}  If removing $D_1$ from the model training
process either reduces or maintains the loss of the resulting
classifier on the poison test data,  \eg $L_{\mathcal{D}_p}(\mathcal{F})
\geq L_{\mathcal{D}_p}(\mathcal{F}^-)$,  this action has skewed the
ratio towards poison data in the training dataset, implying that $D\setminus D_1$ is more responsible for the
success of poison attacks at test time. 

Next, we prove this result in two cases: i) learning from the entire distribution and ii) learning from the empirical distribution. We note that the former is not possible in practice but is useful pedagogically.

\begin{theorem}\label{thm: imbalance_theorem_1}
[Learning from true distribution]
Consider classifiers $\mathfrak{F}_*$ and $\mathfrak{F}_*^-$ that are
trained directly from the true distributions $\mathcal{D}$ and
$\mathcal{D}^-$,  respectively.  We can show that if 
\begin{align}
	L_{\mcD_p}(\mathfrak{F}_*) \geq L_{\mcD_p}(\mathfrak{F}_*^-),
\end{align}
then $\alpha^- \leq \alpha$.
\end{theorem}

\begin{proof}
By definition, 
\begin{align*} 
\mathfrak{F}_*&=\argmin_{\mathfrak{F}}  L_{\mcD}(\mathfrak{F}) \\
&=\argmin_{\mathfrak{F}} \, \alpha L_{\mcD_b}(\mathfrak{F}) + (1-\alpha) L_{\mcD_p}(\mathfrak{F})
\end{align*}
 and 
 \begin{align*}\vspace{-0.1in}
 	\mathfrak{F}_*^- & =\argmin_{\mathfrak{F}}  L_{\mcD^-}(\mathfrak{F}) \\
 	&=\argmin_{\mathfrak{F}} \,  \alpha^- L_{\mcD_b}(\mathfrak{F}) + (1-\alpha^-) L_{\mcD_p}(\mathfrak{F}).
 \end{align*}

The first implies that

\begin{align}
\forall \mathfrak{F}, \ \alpha L_{\mcD_b}(\mathfrak{F}_*)  &+ (1-\alpha) L_{\mcD_p}(\mathfrak{F}_*) \nonumber  \\
& \leq \alpha L_{\mcD_b}(\mathfrak{F}) + (1-\alpha) L_{\mcD_p}(\mathfrak{F}) \label{eq:sub_1} \\ 
\Rightarrow \alpha L_{\mcD_b}(\mathfrak{F}_*) & + (1-\alpha) L_{\mcD_p}(\mathfrak{F}_*) \nonumber  \\
& \leq \alpha L_{\mcD_b}(\mathfrak{F}_*^-) + (1-\alpha) L_{\mcD_p}(\mathfrak{F}_*^-), \label{eq:1}
\end{align}

and the second
\begin{align}
	\forall \mathfrak{F}, \ \alpha^- L_{\mcD_b}(\mathfrak{F}_*^-)  &+ (1-\alpha^-) L_{\mcD_p}(\mathfrak{F}_*^-) \nonumber  \\
	& \leq \alpha^- L_{\mcD_b}(\mathfrak{F}) + (1-\alpha^-) L_{\mcD_p}(\mathfrak{F}) \label{eq:sub_2} \\ 
	\Rightarrow \alpha^- L_{\mcD_b}(\mathfrak{F}_*^-) & + (1-\alpha^-) L_{\mcD_p}(\mathfrak{F}_*^-) \nonumber  \\
	& \leq \alpha^- L_{\mcD_b}(\mathfrak{F}_*) + (1-\alpha^-) L_{\mcD_p}(\mathfrak{F}_*). \label{eq:2}
\end{align}
We multiply Eq. \ref{eq:1} by $\alpha^-$ and Eq. \ref{eq:2} by
$\alpha$, which gives us 
\begin{align}
& \alpha \alpha^- L_{\mcD_b}(\mathfrak{F}_*^-) + \alpha (1-\alpha^-) L_{\mcD_p}(\mathfrak{F}_*^-) -  \alpha (1-\alpha^-) L_{\mcD_p}(\mathfrak{F}_*) \nonumber \\
\leq \, & \alpha \alpha^- L_{\mcD_b}(\mathfrak{F}_*^-) + \alpha^- (1-\alpha) L_{\mcD_p}(\mathfrak{F}_*^-) - \alpha^- (1-\alpha) L_{\mcD_p}(\mathfrak{F}_*) \nonumber \\
\Rightarrow \, &(\alpha-\alpha^-)(L_{\mcD_p}(\mathfrak{F}_*) - L_{\mcD_p}(\mathfrak{F}_*^-)) \geq 0 \label{eq:3}
\end{align}
From Eq. \ref{eq:3}, it is clear that if $L_{\mcD_p}(\mathfrak{F}_*) \geq L_{\mcD_p}(\mathfrak{F}_*^-)$, $\alpha \geq \alpha^-$, proving our claim.
\end{proof}


Our proof did not make any assumptions about the convexity of the loss function or the type of learning algorithm used. This is due to the assumption that we are able to find classifiers that minimize the loss on the true distribution. This assumption does not hold in practice, and we typically use gradient descent algorithms over sampled data for training \cite{shalev-shwartz_understanding_2014}. The next theorem deals with this case, but makes the additional assumptions that the set of possible classifiers is convex and that the loss function is convex, Lipschitz and bounded. The learning algorithm used is Stochastic Gradient Descent (SGD). 

\begin{theorem}\label{thm: imbalance_theorem_2}
[Learning from empirical distribution]
Consider datasets $D$ and $D_2$ defined as above such that $D_2 \subset D$. The corresponding models $\mathfrak{F}$ and $\mathfrak{F}^-$ are defined over a $B$-bounded convex set and trained using SGD over a convex, $\rho$-Lipschitz loss function $\ell$. $D$ is drawn from $\mcD=\alpha \mcD_{b} + (1-\alpha) \mcD_p$, and $D_2$ from $\mcD'=\alpha^- \mcD_{b} + (1-\alpha^-) \mcD_p$. Then, we can show that if 
\begin{align}
	L_{\mcD_p}(\mathcal{F})- L_{\mcD_p}(\mathcal{F}^-) \geq \frac{-(\alpha \epsilon^- + \alpha^- \epsilon)}{\alpha-\alpha^-} \label{eq: thm2}
\end{align}
then $\alpha^- < \alpha$.
\end{theorem}

\begin{proof}
In the learning setting of interest, we have, from Corollary 14.12 in Shalev-Shwartz and Ben-David \cite{shalev-shwartz_understanding_2014},
\begin{align}
	  L_{\mcD}(\mathcal{F})  \leq L_{\mcD}(\mathfrak{F}_*) + \epsilon, \label{eq: convex_1}
\end{align}	
where $\epsilon \geq \frac{B^2 \rho^2}{\vert D \vert}$. Similarly, for learning from $D^-$, we have
\begin{align}
	L_{\mcD^-}(\mathcal{F}^-)  \leq L_{\mcD^-}(\mathfrak{F}^-_*) + \epsilon^-, \label{eq: convex_2}
\end{align}	 
	where $\epsilon^- \geq \frac{B^2 \rho^2}{\vert D^- \vert}$.
	
	Now we can i) substitute $\mathcal{F}^-$ in the right hand side of Eq. \ref{eq:sub_1} and $\mathcal{F}$ in the right hand side of Eq. \ref{eq:sub_2}, ii) use Eqs. \ref{eq: convex_1} and \label{eq:convex_2} and iii) perform the appropriate scaling and rearrangement to get
	\begin{align}
		(\alpha-\alpha^-)(L_{\mcD_p}(\mathcal{F})- L_{\mcD_p}(\mathcal{F}^-) )\geq -(\alpha \epsilon^- + \alpha^- \epsilon).
	\end{align}
	If the condition from Eq. \ref{eq: thm2} is true, then we have $\alpha > \alpha^-$.
\end{proof}

We can then determine that $D_1$ was \emph{less responsible} than the
remaining data $D\setminus D_1$ if the difference of losses satisfies the
condition from Theorem \ref{thm: imbalance_theorem_2}. In other words,
this implies that the remaining data $D\setminus D_1$ is more skewed towards the poison distribution $\mcD_p$, guiding our search for the set of poisoned data. The implication of the theorem above is that set searching is viable since for any identified set, its relative impact on the attack incident can be quantitatively determined. We note that the case when the removed set of data $D_1$ contains only benign data is a special case in the theorem above.

\secspace
\subsection{Further experimental details} \label{appsec: exp_details}

\para{Evaluation Dataset. } We discuss in details of training datasets we used for the evaluation. 
\begin{packed_itemize} \vspace{-0.05in}

\item {\em Image Recognition (CIFAR10)} - The task is to recognize $10$ different objects. The dataset contains 50,000 training images and 10,000 testing images~\cite{krizhevsky2009cifar}. The model is an Wide Residual Neural
Network (RNN) with 50 residual blocks and 1 dense
layer~\cite{zagoruyko2016wide}. We use this task because of its prevalence in general
image classification and security literature.

\item {\em Image Recognition (ImageNet)} - The task is to recognize $1000$ different objects. The dataset contains 1,281,167 training images~\cite{deng2009imagenet}. We include this task because it has been used as a general benchmark for computer vision and the large number of training data poses a challenge for our traceback system. 

\item {\em Face Recognition} (VGGFace) -- This task is to 
  recognize faces of $2,622$ different people drawn from the Internet. We include this task
  because it simulates a more complex facial recognition-based security screening
  scenario. Tracing back poison attack in this setting is
  important. Furthermore, the large set of labels and training data in this task
  allows us to explore the scalability of our system. 

\item {\em Malware Detection} (EMBER Malware) -- Ember is a representative public dataset of malware and goodware samples. The dataset consists of 2,351-dimensional feature vectors extracted from Portable Executable (PE) files for the Microsoft Windows OS. We include the dataset to test traceback performance on malware detection. \vspace{-0.05in}

\end{packed_itemize}


\begin{table}[h]
\centering
\resizebox{0.3\textwidth}{!}{
\begin{tabular}{ccc}
\hline
\begin{tabular}[c]{@{}c@{}}Percentage of \\ Weights Kept\end{tabular} & Precision & Recall \\ \hline
0.1\% & $98.9 \pm 0.0\%$ & $98.0 \pm 0.0\%$ \\
1\% & $99.1 \pm 0.0\%$ & $97.9 \pm 0.1\%$ \\
5\% & $99.2 \pm 0.0\%$ & $97.9 \pm 0.1\%$ \\ \hline
\end{tabular}
}
\caption{Precision and recall of traceback system remain the same as the precentage of weights kept for clustering increases for ImageNet-BadNet.}
\label{tab:p_kept}   \vspace{-0.1in}
\end{table}

\begin{table}[h]
\centering
\resizebox{0.45\textwidth}{!}{
\begin{tabular}{cl}
\hline
\textbf{Attack Method} & \multicolumn{1}{c}{\textbf{Attack Configuration}} \\ \hline
PGD & $\epsilon=0.05$, step size = 9, max iterations = 1000, learning rate = 0.05 \\
CW & $\epsilon=0.05$, \# of iteration = 100, epsilon of each iteration = 0.005 \\
Boundary & $\epsilon=0.05$, num\_iterations = 10000, $\delta = 0.1$ \\
HSJA & $\epsilon=0.05$, num\_iterations = 10000, $\gamma = 1.0$ \\ \hline
\end{tabular}
}
\caption{Detailed information on evasion attacks.}
\label{tab:evasion_details}
\end{table}

\begin{table}[h]
\centering
\resizebox{0.39\textwidth}{!}{
\begin{tabular}{cccc}
\hline
\textbf{\begin{tabular}[c]{@{}c@{}}Perturbation \\ Budget ($L\_inf$)\end{tabular}} & \textbf{\begin{tabular}[c]{@{}c@{}}Attack \\ Success Rate\end{tabular}} & \textbf{Precision} & \textbf{Recall} \\ \hline
0.01 & 29.2\% & 99.4\% & \textbf{94.9\%} \\
0.03 & 86.1\% & 98.4\% & 96.8\% \\
0.05 & 93.7\% & 99.2\% & 99.2\% \\
0.09 & 97.6\% & 99.1\% & \textbf{99.4\%} \\ \hline
\end{tabular}
}
\caption{For \textsf{BP-CIFAR10}, both attack success rate and
  traceback recall increase with the attack perturbation budget,
  because the poison training data becomes more effective.} 
\label{tab:budget_bp}
\end{table}

\begin{table}[t]
\centering
\resizebox{0.45\textwidth}{!}{
\begin{tabular}{lcccc}
\hline
\multicolumn{1}{l}{\textbf{Attack Name}} & \textbf{Dataset} & \textbf{\begin{tabular}[c]{@{}c@{}}Injection \\ Rate\end{tabular}} & \textbf{\begin{tabular}[c]{@{}c@{}}Benign \\ Classification Accuracy\end{tabular}} \\ \hline
BadNet & CIFAR10 & $10\%$ & $92.9 \pm 0.3\%$ \\
BadNet & ImageNet & $10\%$ & $78.9 \pm 1.7\%$ \\
Trojan & VGGFace &  $10\%$ & $76.1 \pm 0.8\%$ \\
Physical Backdoor & Wenger Face & $10\%$ & $99.9 \pm 0.0\%$ \\ \hline
\end{tabular}
}
\caption{\htedit{The default setup of dirty-label poisoning attacks.}}
\label{tab:attack_detail}   
\end{table}

\begin{table}[t]
\centering
\resizebox{0.45\textwidth}{!}{
\begin{tabular}{lcccc}
\hline
\multicolumn{1}{c}{\textbf{Attack Name}} & \textbf{Dataset} & \textbf{\begin{tabular}[c]{@{}c@{}}Injection \\ Rate\end{tabular}} & \textbf{\begin{tabular}[c]{@{}c@{}}Benign \\ Classification Accuracy\end{tabular}}  \\ \hline
BP & CIFAR10 & $0.01\%$ & $93.0 \pm 0.2\%$ \\
BP & ImageNet & $0.01\%$ & $79.1 \pm 0.9\%$ \\
WitchBrew & CIFAR10 & $1\%$ & $92.2 \pm 0.3\%$ \\
WitchBrew & ImageNet & $1\%$ & $79.3 \pm 0.7\%$ \\ 
Malware Backdoor & EMBER Malware & $1\%$ & $99.2 \pm 0.1\%$  \\ \hline
\end{tabular}
}
\vspace{-0.1in}
\caption{The default setup of the five clean-label poisoning attacks
  used in our evaluation.} \vspace{-0.1in}
\label{tab:clean_label_detail}
\end{table}

\begin{table}[h]
\centering
\resizebox{0.45\textwidth}{!}{
\begin{tabular}{lccc}
\hline
\multirow{2}{*}{\textbf{Attack-Dataset}} & \multicolumn{2}{c}{\textbf{Avg
                                        $L_2$ distance of poison data
                                        to}} &
                                               \multirow{2}{*}{\textbf{\begin{tabular}[c]{@{}c@{}}Avg \# of\\ iterations\end{tabular}}} \\ \cline{2-3}
 & \textbf{benign centroid} & \textbf{poison centroid} &  \\ \hline
BadNet-CIFAR10 & 0.65 & 0.19 & 3.2 \\
BadNet-ImageNet & 0.76 & 0.24 & 3.4 \\
Trojan-VGGFace & 0.69 & 0.09 & 3.1 \\
Physical-Wenger & 0.62 & 0.39 & 4.0 \\ \hline
\end{tabular}
}
\caption{The average $L_2$ distance between individual poison data and
  the benign (poison) centroid, and
  the number of pruning iterations needed to complete the traceback. }
\label{tab:dist_cluster}
\end{table}

\begin{figure}[t]
  \centering
    \includegraphics[width=0.35\textwidth]{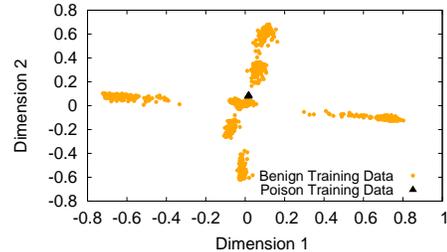}
    \vspace{-0.1in}
    \caption{2-D PCA visualization of the projection of training data (sampled from \textsf{BP-CIFAR10}). Orange circles are innocent data and red crosses are poison data. }
    \label{fig:cluster_bp_pca}
  \end{figure}

\end{document}